\documentclass[twoside,12pt]{article}
\usepackage[OT1,T2A]{fontenc}
\usepackage[cp1251]{inputenc}

\usepackage[russian]{babel}
\usepackage{epsfig}
\usepackage{amsmath,amsfonts}
\usepackage{graphicx}
\usepackage{amssymb}
\usepackage{amsxtra}
\usepackage{amsthm}
\usepackage[mathscr]{eucal}
\usepackage{tabularx}
\usepackage{indentfirst}
\usepackage{float}

\newcommand\RE{\operatorname{Re}}

\begin{document}
\renewcommand{\refname}{{\large \bf References}}
\renewcommand{\figurename}{Figure }
\renewcommand{\tablename}{Table }

\vspace{-4mm}
\begin{center}
\large{\bf Low-order models of 2D fluid flow in annulus }
\end{center}

\markboth{\hfill N.\,V.\,Petrovskaya, M.\,Yu.\,Zhukov  \hfill }
{\hfill Low-order models of 2D fluid flow in annulus \hfill}

\vspace{-4mm}
\begin{center}
\large{ N.\,V.\,Petrovskaya, M.\,Yu.\,Zhukov}

\vspace{2mm}
 \normalsize{\it Department of Mathematics, Mechanics and Computer Science,}  \\
 \normalsize{\it Southern Federal University, 344090, Rostov-on-Don, Russia
 } \\
\end{center}

\vspace{2mm} \small

The two-dimensional flow of viscous incompressible fluid in the
domain between two concentric circles is investigated numerically.
To solve the problem, the low-order Galerkin models are used. When
the inner circle rotates fast enough, two axially asymmetric flow
regimes are observed. Both regimes are the stationary flows
precessing in azimuthal direction. First flow represents the region
of concentrated vorticity. Another flow is the jet-like structure
similar to one discovered earlier in experiments
\cite{Vladimirov-1,Vladimirov-2}.

\vspace{-2mm} \normalsize
\section*{Introduction}

 The interest to the problem considered is stimulated by
the strange flow regimes induced in the thin fluid layer by the
rotating cylinder, experimentally discovered by V.\,A.\,Vladimirov
\cite{Vladimirov-1,Vladimirov-2} in 1994. The flow in a thin annular
gap restricted in axial direction by rigid walls was considered.
When the inner cylinder rotates fast enough, axially asymmetric,
slow-precessing, stable jet-like pattern appears. These jets
represent the sectors with the strong radial outflow which precesses
in azimuthal direction.

The flow regimes observed in experiments are not deeply investigated
and the mechanism of their onset is not clear. The flow observed is
not the Moffatt vortex that appears in the domain with corners
\cite{Moffatt}. It is not the Hammel flow with the azimuthal jets,
investigated and classified by M.\,A.\,Goldshtik, V.\,N.\,Shtern et
al. (see for example \cite{Goldshtik-1,Goldshtik-2,Shtern} where the
flow regimes without the inner cylinder but with the source are
considered). It is not the spiral structure described in works by
M.\,V.\,Nezlin and E.\,N.\,Snezhkin \cite{Nezlin-1,Nezlin-2} that
appears in the thin layer of rotating fluid with a free surface with
differentially rotating parabolic-shape bottom. We notice that in
spite of visual similarity with regimes from
\cite{Nezlin-1,Nezlin-2}, flows in \cite{Vladimirov-1,Vladimirov-2}
are totally different. In \cite{Nezlin-1,Nezlin-2} the reason for
the spiral structure onset is the parabolic bottom profile that
models the gravity field in the radial direction, so that the
phenomenon could be described with the use of shallow water theory.
In other words the jet-like structures in \cite{Nezlin-1,Nezlin-2}
are similar to waves on the free surface. Here it is worth
mentioning works by V.\,Yu.\,Liapidevsky (see for example
\cite{Liapidevsky}) on the two-dimensional vortical shallow water
flows in a gap between two rigid boundaries.

There is a number of works on the similar problems. These works are
mainly devoted to study the axially symmetric regimes in the
Couette-Taylor flow between short cylinders. These problems were
intensively investigated numerically after T.\,B.\,Benjamin and
T.\,Mullin \cite{Benjamin}. The axially symmetric flows in the
annular gap with comparable radial and axial sizes, governed by the
two-dimensional Navier-Stokes equations were studied with the help
of the finite-difference methods (particularly the methods of
markers and cells, and their modifications) and of the Galerkin
method (see for example \cite{Benjamin, Cliffe, Lucke, Pfister,
Kobine}). The base axially symmetric flow regime in the domain with
two rigid end-walls represents two Taylor vortices. The bifurcation
of this flow to the regime with one major vortex and one minor
vortex is described.

In works \cite{Vladimirov-3, Vladimirov-4, Vladimirov-5,
Vladimirov-6} the asymptotic model describing the base axially
symmetric regime in the thin in axial direction annular domain
between two cylinders is constructed and compared with experiment.
This asymptotic model describes two Taylor vortices in the case of
rigid end-walls and one Taylor vortex in the case when one of the
end-walls is the non-deformable free surface. Particularly, the
analytic formula has been obtained for the azimuthal velocity (it
was earlier defined numerically in \cite{Cliffe, Lucke, Pfister,
Kobine}). The azimuthal velocity calculated with the help of this
analytic formula is in a good agreement with the experimental data
\cite{Vladimirov-3, Vladimirov-4, Vladimirov-5, Vladimirov-6}.

In the work presented we attempt to model some of the experimental
results from \cite{Vladimirov-1, Vladimirov-2} with the use of the
low-order Galerkin model, applied to the viscous flow in the
two-dimensional annular strip. When the base flow similar to the
Couette flow loses stability for the low-order model, the stationary
asymmetric flow (the soliton-like vortex) appears. This vortex
slowly precesses in the azimuthal direction. Furthermore, numerical
experiments with the low-order models allowed to discover another
axially asymmetric stationary flow regime: slow precessing in the
azimuthal direction jet-like structure. This axially asymmetric
solution (apart from the first one) does not bifurcate from the base
flow but appears 'from the sky-blue'. The jet-like regimes appear to
be similar to ones observed in experiments presented in
\cite{Vladimirov-1,Vladimirov-2}. It is rather surprising because,
according to \cite{Vladimirov-1,Vladimirov-2, Vladimirov-3,
Vladimirov-4,Vladimirov-5,Vladimirov-6}, the jet-like flows observed
in the narrow (in axial direction) gap are substantially
three-dimensional. Another surprise is that the Reynolds numbers for
which asymmetric structures appear are of the same order as in the
experiments. Comparison of the other parameters is impossible since
the numerically solved problem is two-dimensional while the
experimental facility is essentially three-dimensional.

The certain uneasiness is caused by dependence of the results of
numerical modeling on the number of radial basis functions for
Galerkin model. With the increase of the number of this functions,
increase of the critical Reynolds numbers is observed (note that the
results weakly depend on the number of azimuthal basis functions).
In other words, results obtained (existence of the precessing
stationary flows in the form of the soliton-like jet or the
soliton-like vortex) could be specific for the low-order Galerkin
models only, while the base flow could be stable for the high-order
models. However, good qualitative agreement with the experiments
described in \cite{Vladimirov-1,Vladimirov-2} make us hope that the
low-order models are useful to improve the understanding of the
mechanism of jet-like structures onset in the gaps narrow in the
axial direction.

\setcounter{equation}{0} \setcounter{section}{0}
\section{Basic equations}

Let the annular strip domain bounded by the circles of radii $r_1$ и
$r_2$ ($r_1<r_2$) is filled by the viscous incompressible fluid. The
inner boundary $r=r_1$ is rotating with the constant azimuthal
velocity $\Omega$, and the outer boundary $r=r_2$ is fixed. The
Navier-Stokes equations in the polar coordinates  $(r,\theta)$ in
the dimensionless form are:
\begin{eqnarray}
&&  u_t+ u u_r + \frac{1}{r} \,\, v u_\theta -
 \frac{v^2}{r}= -p_r+
 \RE^{-1} \left(\Delta u-\frac{u}{r^2}-\frac{2}{r^2} \,\,
 v_\theta\right),
 \nonumber \\
&&  v_t+ u v_r + \frac{1}{r} \,\, v v_\theta +
 \frac{uv}{r}= -\frac{1}{r} \,\, p_\theta+
 \RE^{-1} \left(\Delta v-\frac{v}{r^2}+\frac{2}{r^2} \,\,
 u_\theta\right),
 \label{1.1}\\
&&  (r u)_r + v_\theta =0, \nonumber
\end{eqnarray}
\begin{equation}
\Delta (\quad)=(\quad)_{rr} + \frac{1}{r} \,\, (\quad)_r +
\frac{1}{r^2} \,\,(\quad)_{\theta\theta}. \nonumber
\end{equation}

Here $u$ is the radial velocity, $v$ is the azimuthal velocity, $p$
is the pressure, $\RE$ is the Reynolds number.

Boundary conditions at $r=1$ and $r=b$ have the form:
\begin{equation}
\left.  u\right|_{r=1}=0, \quad \left. v\right|_{r=1}=1, \quad
\left. u\right|_{r=b}=0,\quad  \left. v\right|_{r=b}=0. \label{1.2}
\end{equation}

Dimensionless variables $t$, $r$, $u$, $v$, $p$ correspond to
dimensional variables $t'$, $r'$, $u'$, $v'$, $p'$:
\begin{eqnarray}
&& t'=\frac{t}{\Omega},\quad  r'=rr_1,\quad
 (u',v')=\Omega r_1(u,v),\quad  p'=p\rho \Omega^2 r_1^2, \nonumber \\
&& \RE=\frac{\Omega r_1^2}{\nu},  \quad  \varepsilon=\RE^{-1}, \quad
b=\frac{r_2}{r_1}. \label{1.3}
\end{eqnarray}
where $\nu$ is the kinematic viscosity, $r_1$ is the radius of the
inner circle, $r_2$ is the radius of the outer circle, $\Omega$ is
the azimuthal velocity of the inner circle rotation, $\rho$ is the
constant fluid density.

The problem (\ref{1.1})--(\ref{1.2}) has the stationary axially
symmetric solution (Couette flow):
\begin{eqnarray}
 u=0,\quad v=V_0(r),\quad V_0(r)=\frac\alpha r+\beta r,\quad
 \alpha=\frac{b^2}{b^2-1},\quad  \beta=-\frac1{b^2-1}.
\label{1.4}
\end{eqnarray}

Let $\Psi(t,r,\theta)$ be a stream function  and $\psi(t,r,\theta)$
be a perturbation of the stream function:
\begin{equation}
  u=-\frac1r\psi_\theta=-\frac1r\Psi_\theta,\quad  v=\psi_r+V_0=\Psi_r.
\label{1.5}
\end{equation}

If we substitute (\ref{1.5}) to (\ref{1.1})--(\ref{1.2}), we obtain:
\begin{eqnarray}
&&  \Delta \psi_t+\frac1r\bigl(\psi_r(\Delta\psi)_\theta-
 \psi_\theta(\Delta\psi)_r\bigr)=\varepsilon
\Delta^2\psi-\frac{V_0}{r}(\Delta\psi)_\theta, \label{1.6} \\ [2mm]
&& \left.  \psi\right|_{r=1}=0, \quad \left. \psi_r\right|_{r=1}=0,
\quad \left. \psi\right|_{r=b}=0,\quad  \left.
\psi_r\right|_{r=b}=0. \quad \label{1.7}
\end{eqnarray}

\setcounter{equation}{0}
\section{Galerkin approximations}

To obtain the approximate solution of the problem
(\ref{1.6})--(\ref{1.7}), Galerkin method is used. We approximate
the stream function perturbation $\psi$ as follows:
\begin{equation}\label{2.1}
  \psi=2\sum_{m=1}^L\sum_{n=1}^N
  \left\{(x_{m n}(t)\cos(n\theta)-y_{m n}(t)\sin(n\theta)
  \right\}f_{m n}(r).
\end{equation}
Here, the functions $f_{m n}(r)$  satisfy the boundary conditions:
\begin{equation}\label{2.2}
 f_{m n}(1)=0,\quad f'_{m n}(1)=0,\quad f_{m n}(b)=0,\quad f'_{m n}(b)=0,\quad
\end{equation}

The standard Galerkin procedure reduces the problem
(\ref{1.6})--(\ref{1.7}) to the system of the ordinary differential
equations for Galerkin coefficients  $x_{m n}(t)$, $y_{m n}(t)$. We
also assume that the functions $f_{m n}(r)$ satisfy the following
conditions:
\begin{equation}\label{2.3}
\int\limits_1^b f_{m n}(r)\Delta_n f_{jn}(r) r\,dr=-\delta_{jm}
\end{equation}
\begin{equation}\label{2.4}
\Delta_n(\,\,)=(\ )_{rr}+\frac1r(\ )_r-\frac{n^2}{r^2}(\ ).
\end{equation}

In the work  \cite{Ponomarev} the two-dimensional flow in the
annular strip (driven by the azimuthal force changing the sign with
radius) is studied. Analogously to the work \cite{Ponomarev} the
functions  $f_{m n}(r)$ for fixed $n$ are taken as the
eigenfunctions of the following boundary value problem:
\begin{eqnarray}
&&  \Delta_n^2 f + h^2 \Delta_n f=0, \label{2.5} \\ [2mm] &&
 f(1)=0,\quad f'(1)=0,\quad f(b)=0,\quad f'(b)=0 \label{2.6}
\end{eqnarray}

Calculation of the eigenvalues  $h_{m n}$ for the problem
(\ref{2.5})--(\ref{2.6}) and of the corresponding eigenfunctions
$f_{mn}(r)$ is a tiresome procedure. That is why we also use the
polynomial basis to construct functions  $f_{m n}(r)$.

To determine Galerkin coefficients $x_{m n}(t)$, $y_{m n}(t)$, the
following system of the ordinary differential equations is obtained:
\begin{equation}
 \frac{dx_{m n}}{dt} = -\varepsilon \sum_{j=1}^L \delta_{m jn}x_{jn}
  - \sum_{j=1}^L \gamma_{m jn}y_{jn} -\sum_{j=1}^L\sum_{p=1}^L\sum_{k=1}^N\sum_{q=1}^N
  (\alpha_{kqn}x_{jk}y_{pq}-\beta_{kqn}y_{jk}x_{pq}) K_{jkpq m n},
 \label{2.7}
\end{equation}
\begin{equation}
 \frac{dy_{m n}}{dt} = -\varepsilon \sum_{j=1}^L \delta_{m jn}y_{jn}
  + \sum_{j=1}^L \gamma_{m jn}x_{jn}+ \sum_{j=1}^L\sum_{p=1}^L\sum_{k=1}^N\sum_{q=1}^N
  (\beta_{qnk}x_{jk}x_{pq}-\beta_{knq}y_{jk}y_{pq})
  K_{jkpq m n}. \nonumber
\end{equation}
Here
\begin{equation}
\alpha_{kqn}=1, \  \text{if\ }\  n=|k-q| \ \ \text{or\ }\ n=k+q; \ \
\alpha_{kqn}=0, \ \text{otherwise}; \nonumber
\end{equation}
\begin{equation}
\beta_{kqn}=1, \ \text{if\ }\ n=|k-q|; \ \ \beta_{kqn}=-1, \
\text{if\ }\ n=k+q; \ \ \beta_{kqn}= 0, \ \text{otherwise}.
\nonumber
\end{equation}
The coefficients $\delta_{m jn}$, $\gamma_{m jn}$ and $K_{jkpq m n}$
are defined by the following relations:
\begin{eqnarray}
&& \delta_{m jn}=\int\limits_1^b f_{m n}\Delta_n g_{jn}
  r\,dr, \quad \gamma_{m jn}=n\int\limits_1^b V_0 f_{m
  n}g_{jn} \,dr,
  \label{2.8}\\ [1mm]
&& K_{jkpq m n} =q \int\limits_1^b (f'_{jk}g_{pq}-g'_{jk}f_{pq})
f_{m n}\,dr, \quad g_{jn}=\Delta_n f_{jn}. \nonumber
\end{eqnarray}

\setcounter{equation}{0}
\section{Numerical experiments}

Solutions of the system of equations (\ref{2.7}) depend on the two
dimensionless parameters: the Reynolds number $\RE$ (or
 $\varepsilon=\RE^{-1}$)  and on the ratio of the outer and inner radii
$b$. Note that the parameter $\varepsilon$ is included to the system
(\ref{2.7}) explicitly whereas the coefficients $\delta_{m jn}$,
$\gamma_{m jn}$ и $K_{jkpq m n}$ depend on the parameter $b$.

\subsection{Basis functions}

Two sets of the basis functions are in use: the set of solutions of
the problem (\ref{2.5})--(\ref{2.6}), and the linear combinations of
the polynomials $P_m(r)=(r-1)^2(b-r)^2r^{m-1}$ which satisfy the
conditions (\ref{2.2})--(\ref{2.4}).

Solutions of the problem (\ref{2.5})--(\ref{2.6}) have the form
\cite{Ponomarev}:
\begin{equation}\label{pr1}
  f_{m n}(r)=\bigl[A_{m n} J_n(\rho)+
  B_{m n} Y_n(\rho)+C_{m n}\rho^n +D_{m n} \rho^{-n}
   \bigr]/ h_{m n}^2,\,\,\rho=h_{m n}r \nonumber
\end{equation}
where $J_n$ is the Bessel function of the first kind, $Y_n$ is the
Neumann function.

The eigenvalues $h_{m n}$ and coefficients $A_{m n}$, $B_{m n}$,
$C_{m n}$, $D_{m n}$ are defined by the boundary conditions:
\begin{equation}\label{pr2}
  f_{m n}(1)=0,\quad f'_{m n}(1)=0,\quad
  f_{m n}(b)=0,\quad f'_{m n}(b)=0
\end{equation}

Values $h_{m n}$, ($m=1,2,\dots,L$) are the roots of the determinant
of the system (\ref{pr2}). The nontrivial solution $(A_{m n}, B_{m
n}, C_{m n l}, D_{m n})$ of the system (\ref{pr2}) must satisfy the
condition (\ref{2.3}). The function $ g_{m n}$ has the form:
\begin{equation}\label{pr3}
 g_{m n}=\Delta_n f_{m n}=A_{m n} J_n(\rho)+ B_{m n} Y_n(\rho). \nonumber
\end{equation}

The system of functions $P_m(r)=(r-1)^2(b-r)^2r^{m-1}$ is used to
construct the polynomial basis. Functions $f_{m n}(r)$ for fixed $n$
are constructed with the help of Gramm-Shchmidt orthogonalization
with the scalar product:
\begin{equation}\label{pr4}
 (u,v)=-\int_1^b u(r)\Delta_n v(r)r\,dr, \quad (f_{m n},f_{m n})=1. \nonumber
\end{equation}

Chosen sets of basis functions are almost identical. Indeed, as the
numerical experiment showed, properties of the solution of
(\ref{1.6})--(\ref{1.7}) obtained by Galerkin method weakly depend
on the choice of system of the basis functions $f_{m n}(r)$. The
number $L$ of the radial basis functions is chosen to be 2 or 3, and
the number $N$ of the azimuthal basis functions is not more then 9.

\subsection{Dependence of the critical Reynolds numbers on the number of
basis functions}

\begin{figure}[H]
\hspace{3mm}
\includegraphics[width=5.2cm,angle=-90]{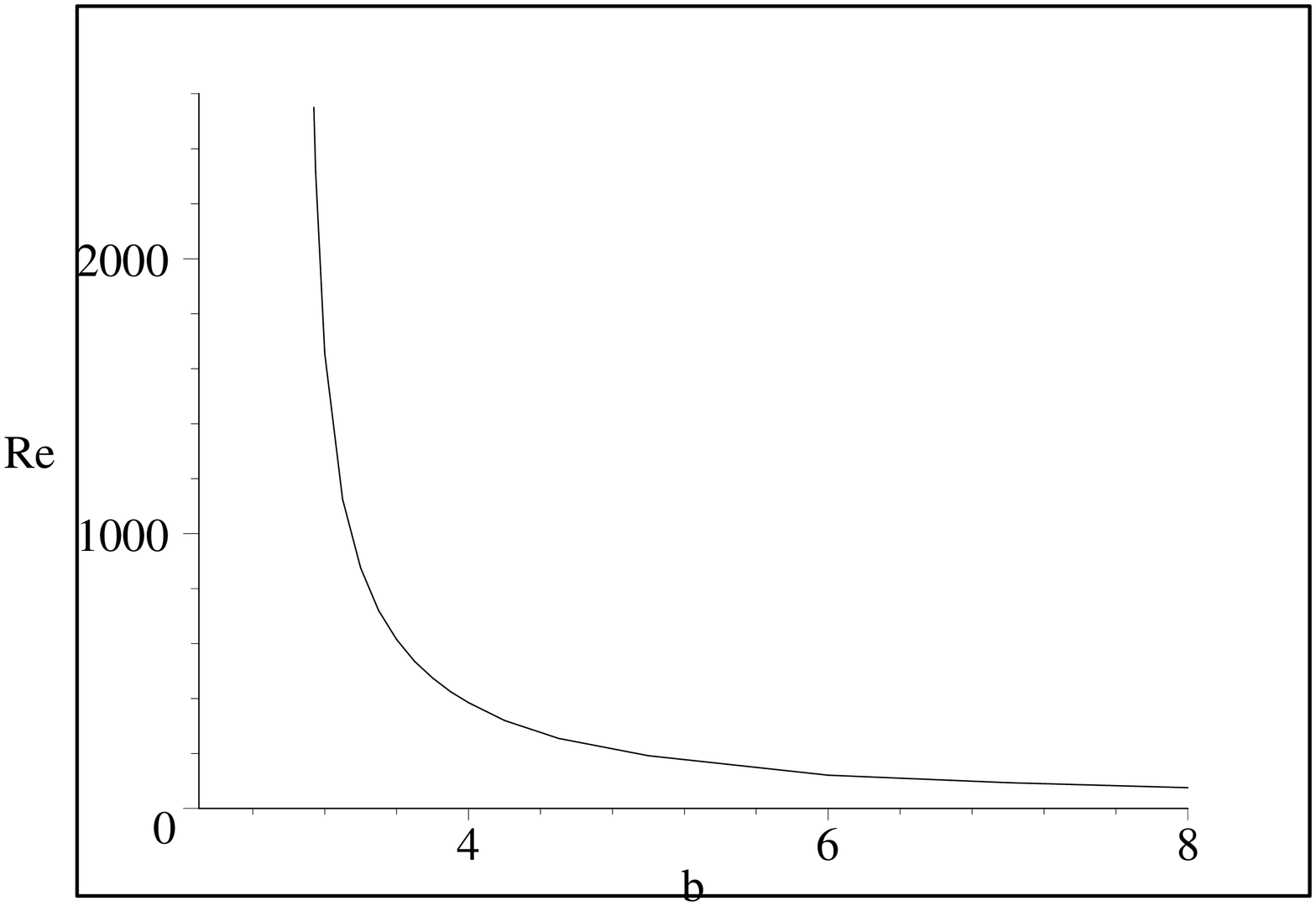}\hspace{6mm}
\includegraphics[width=5.2cm,angle=-90]{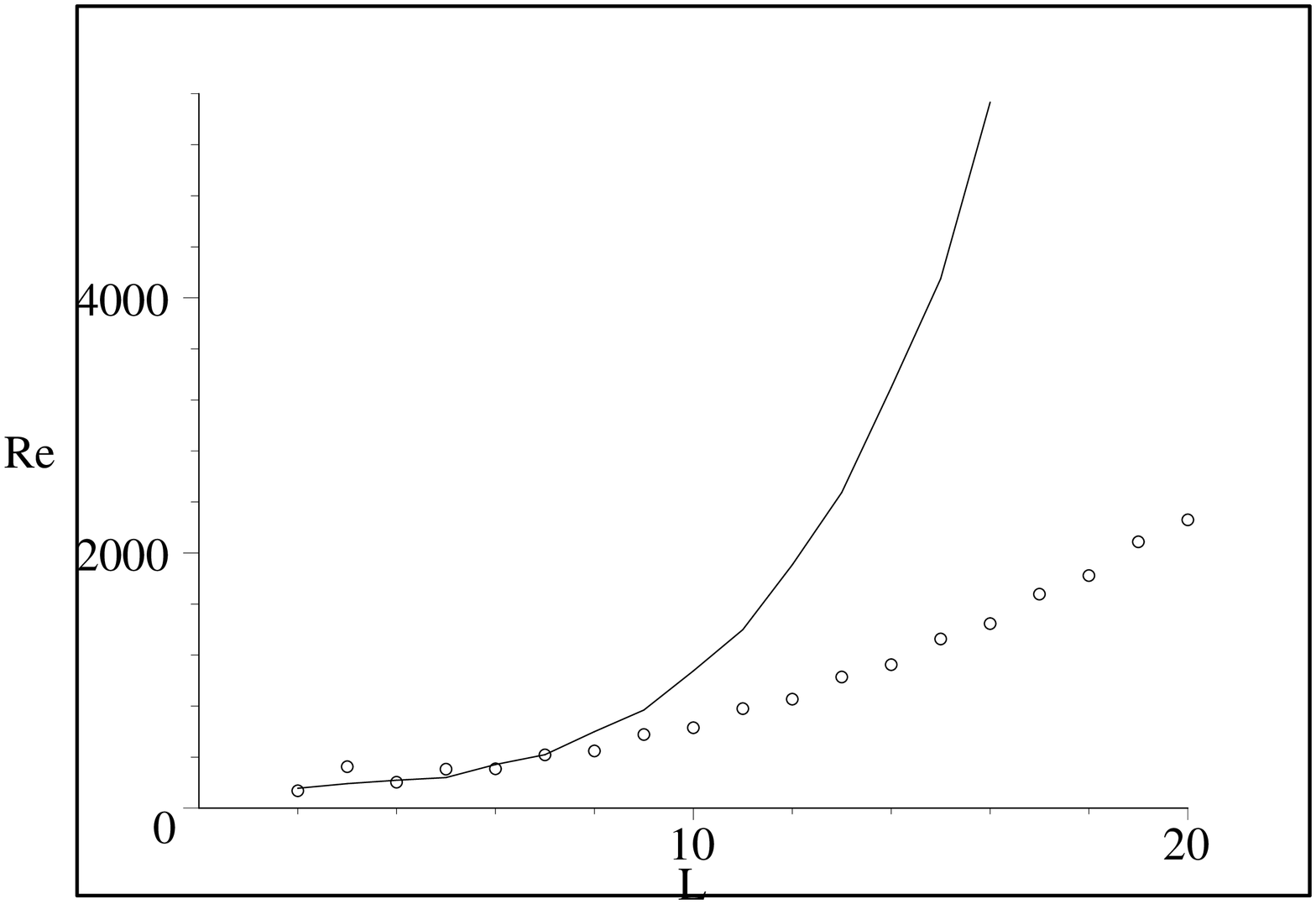}
\caption{Left: the neutral curve $\RE=\RE_0(b)$ of the Couette flow
instability, corresponding to the azimuthal wave number $n=1$
($L=3$, $N=9$, polynomial basis). Right: dependence of the critical
Reynolds numbers on the number of basis functions $L$ at $b=5$.
Continuous line corresponds to computation with polynomial basis;
circles correspond to basis composed of the eigenfunctions of
(\ref{2.5})--(\ref{2.6}); azimuthal wave number $n=1$.}
 \label{fig1}
\end{figure}

Equations (\ref{2.7}) have the solution which corresponds to the
base regime (Couette flow). For the small Reynolds numbers this flow
is stable, however, it can become unstable when the Reynolds number
increases. The neutral curve $\RE=\RE_0(b)$ for $N=9$, $L=3$  is
shown in Fig.~\ref{fig1} (left). The neutral curve corresponds to
the oscillatory instability with the azimuthal wave number $n=1$.
Neutral curves with the azimuthal wave numbers $n=2,\,3$ have
similar form.

When the parameter $b$ is fixed, the critical Reynolds numbers
$\RE_0(b)$ are strongly dependent on $L$ (the number of radial basis
functions). Numerical investigation of the linear stability of the
solution $x_{mn}=0$, $y_{mn}=0$ of equations (\ref{2.7}) shows that
the critical Reynolds numbers are monotonously increasing with the
increase of $L$ when $L\geqslant 6$. Results of calculation at $b=5$
are presented in Fig.~\ref{fig1} (right). Continuous line
corresponds to the polynomial basis, and circles represent basis
composed from the eigenfunctions of the problem
(\ref{2.5})--(\ref{2.6}).

\subsection{Flow regimes}

Two different stable stationary regimes (traveling waves) are found
apart from the base Couette flow in numerical experiments performed.
Galerkin coefficients $x_{mn}(t)$ and $y_{mn}(t)$ are periodic
functions of time $t$, while the amplitudes
$d_{mn}=(x_{mn}^2+y_{mn}^2)^{1/2}$ are constant for these regimes.
One of them appears as a result of instability of the base Couette
flow. Nature of the second regime is still not clear. Both flows
have a number of typical features that are independent of the outer
radius $b$, the choice of the basis functions $f_{m n}(r)$ and the
number of these functions $L$ (for small enough $L$) provided that a
number of the azimuthal basis functions is chosen to be large
enough.
\begin{figure}[H] \hspace{7mm}
\includegraphics[width=5.2cm,angle=-90]{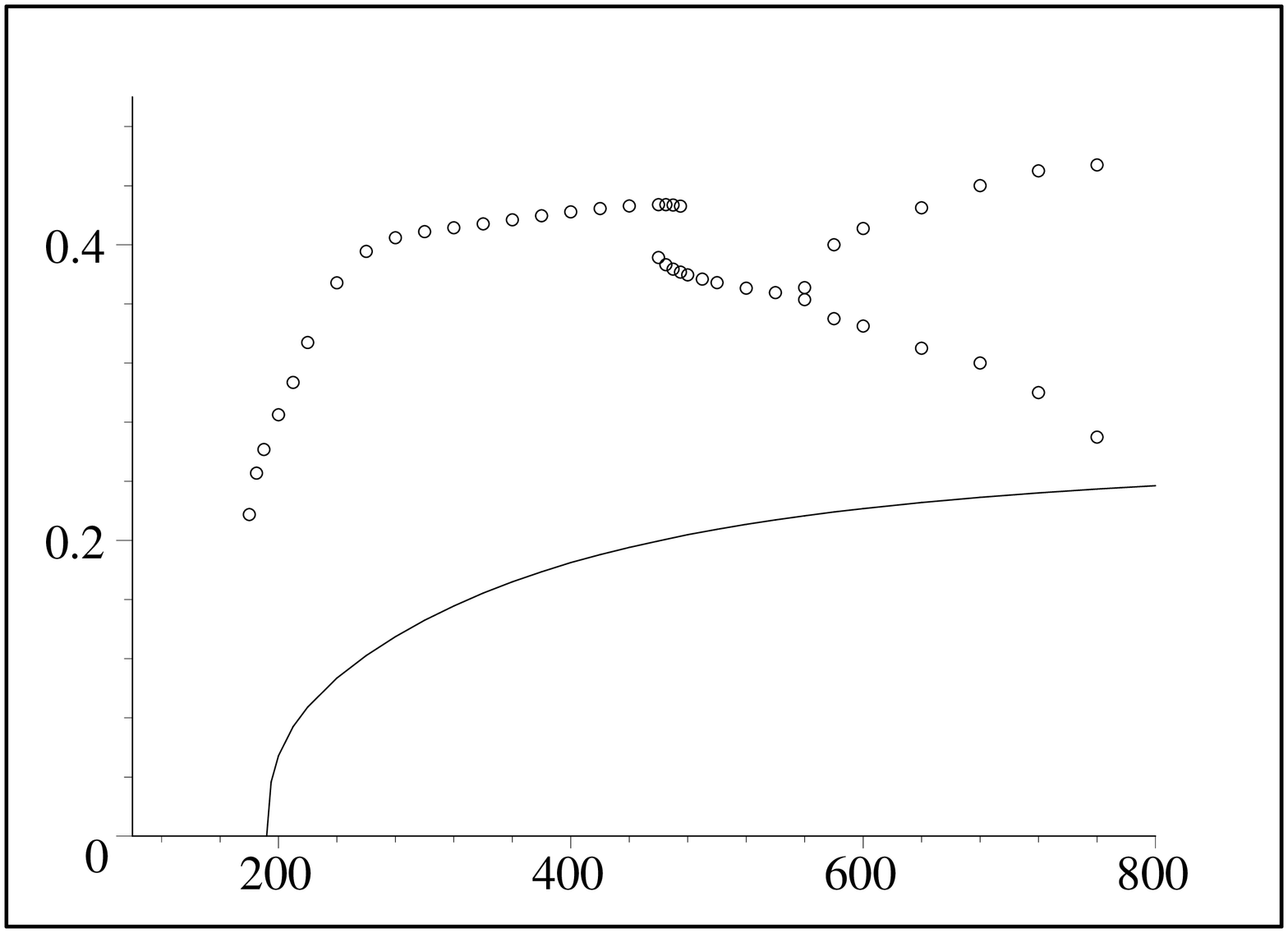}\hspace{8mm}
\includegraphics[width=5.2cm,angle=-90]{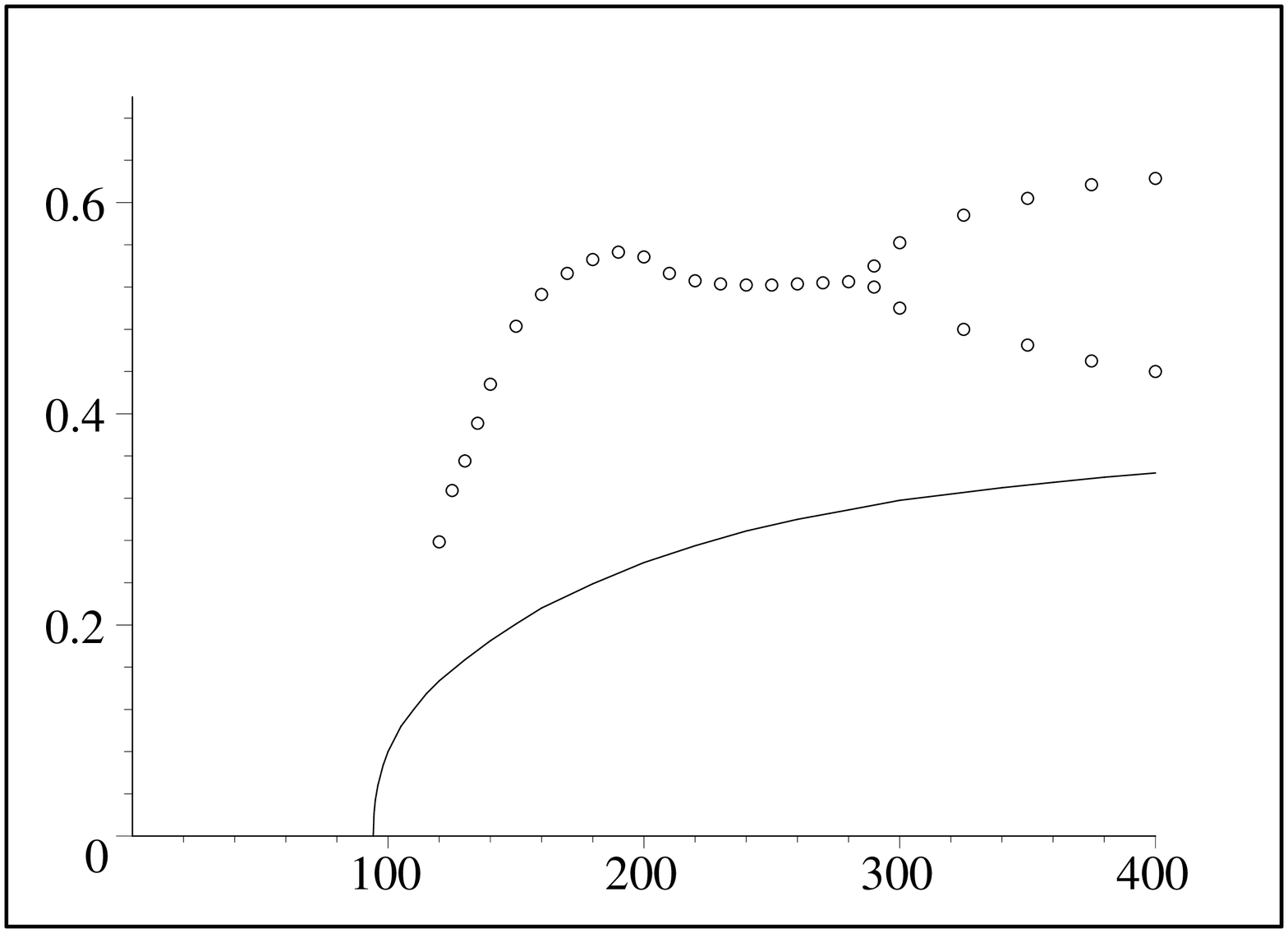}
\caption{Dependence of the amplitude $d_{11}$ on the Reynolds
number: continuous line corresponds to regime with precessing
vortex; circles correspond to regime with the jet. Left picture
correspond to $b=5$, right picture corresponds to $b=7$, polynomial
basis is used in both cases. Regime J is stable at $110\,<\RE <
\,290$ for $b=7$, and at $175\,<\RE < \,480$ for $b=5$.}
\label{fig2}
\end{figure}

In the wide interval of parameters regimes observed are stable. The
onset of one or another of these regimes depends on the initial
conditions. One regime is the stationary flow with the single
precessing vortex spot. We call it 'regime  V'. Main feature of the
second flow regime is the region of intensive flow in the form of
the radial jet, precessing in the azimuthal direction. We call it
'regime J'.

In the models constructed with the use of polynomial basis by  $r$,
the system loses stability smoothly in the sense that the periodical
motion branches to the supercritical area $\RE>\RE_0$ and is stable.
If we use the basis composed from the eigenfunctions of the problem
(\ref{2.5})--(\ref{2.6}), we observe the rough loss of stability:
the unstable limiting cycle branches to the subcritical area.

\begin{figure}[H] \hspace{15mm}
\includegraphics[width=4.6cm,angle=-90]{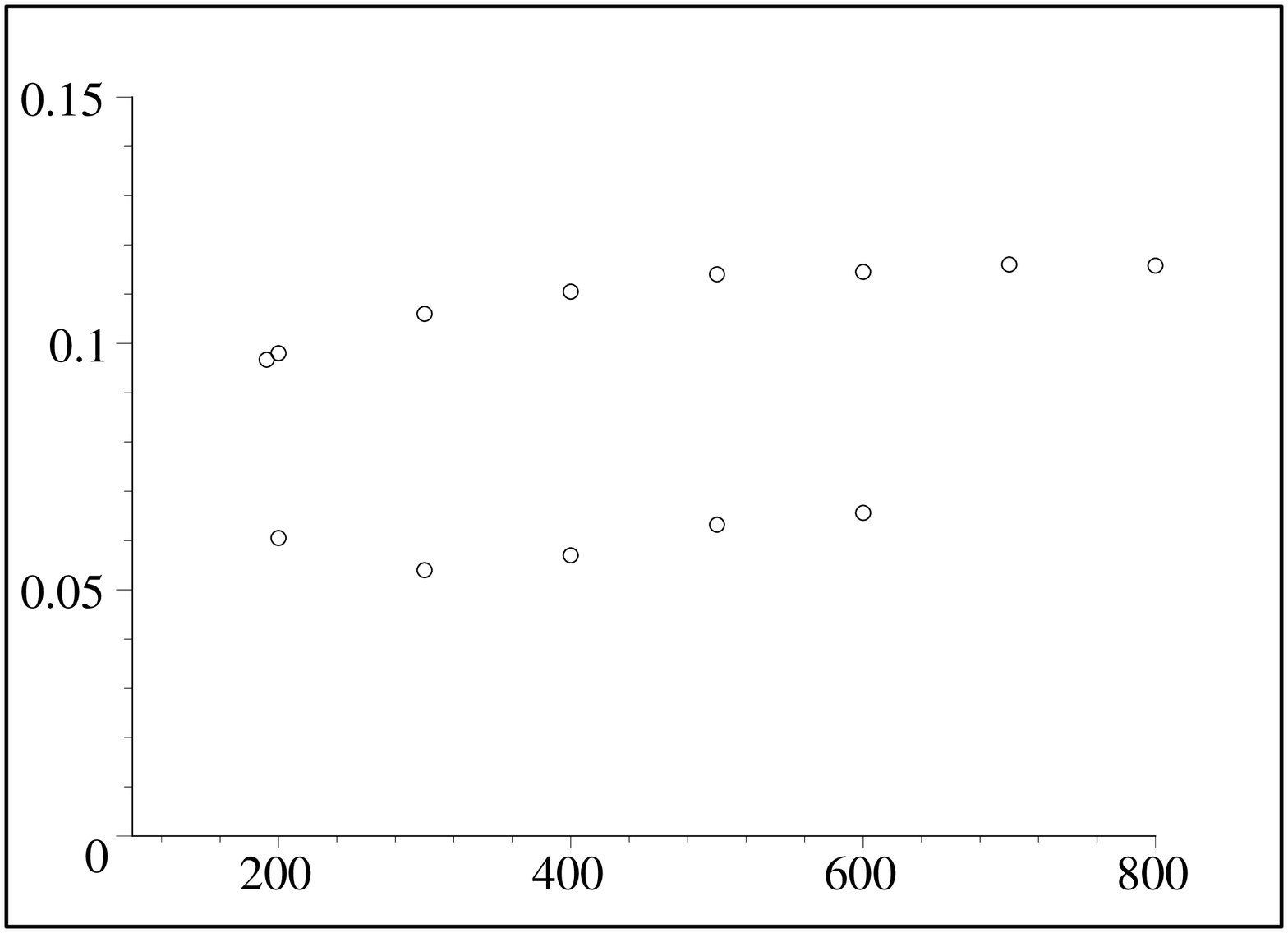} \hspace{8mm}
\includegraphics[width=4.6cm,angle=-90]{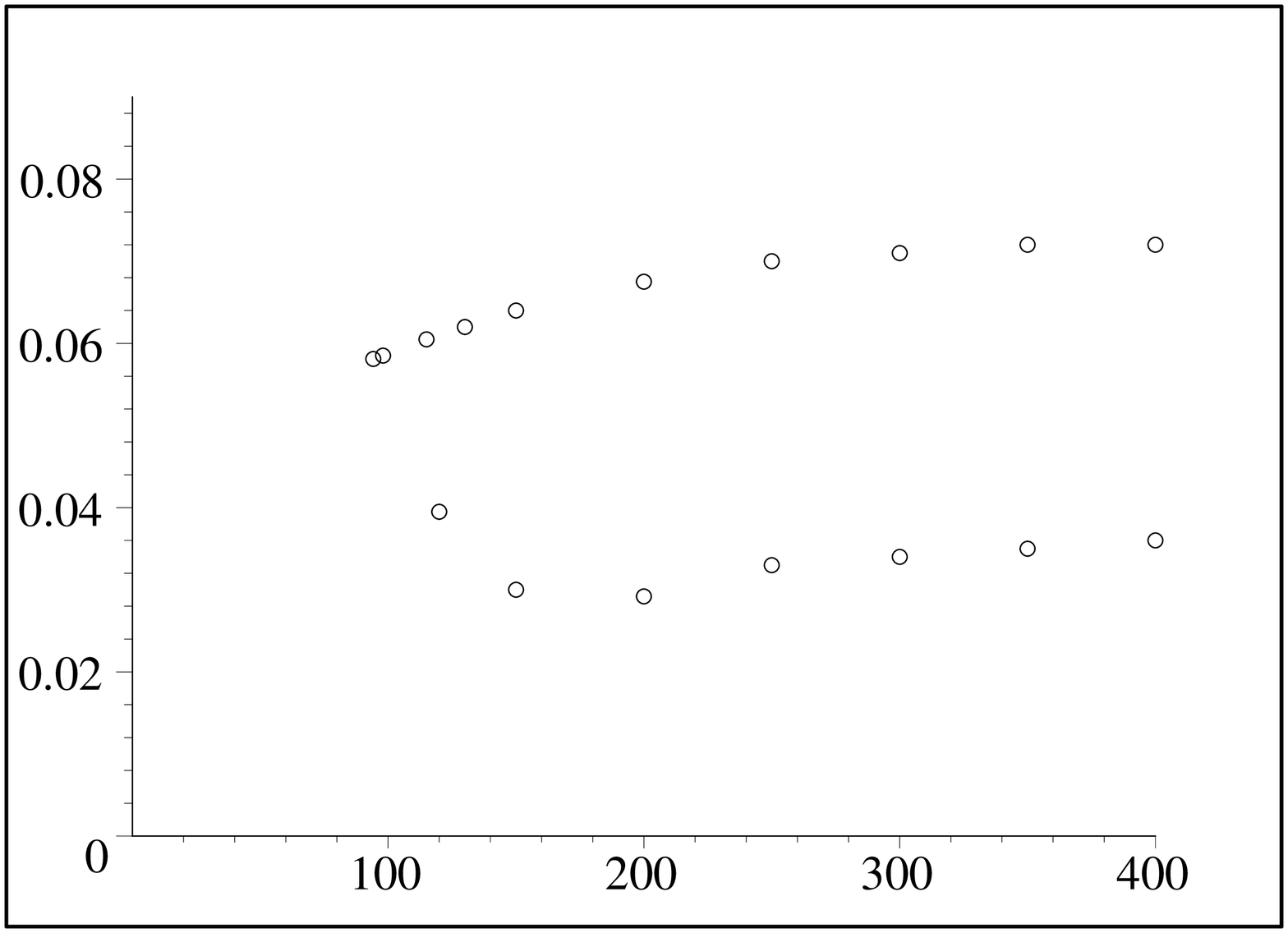}
\vspace{1mm} \caption{Dependence of precession speed on the Reynolds
number: upper points correspond to regime with precessing vortex,
lower points correspond to regime with the jet. Left picture
corresponds to $b=5$, right picture corresponds to  $b=7$.
Polynomial basis is used in both cases.}
 \label{fig3}
\end{figure}

We emphasize again that there is a certain range of parameters where
both stationary flow regimes are stable. The onset of one or another
flow regime depends on initial conditions. Regime V typically
appears when the initial values of $|\,\bf{v}|$ are small. Regime J
is usually observed when the initial values of $\,|\bf{v}|$  are
large. Note that the values of $\,d_{mn}$ that correspond to regime
J are significantly larger than ones for the flow of V type. The
amplitude curves $\,d_{11}=d_{11}(\RE)$ are presented in
Fig.~\ref{fig2}. Note that regime V branches from the base Couette
flow whereas regime J appears 'from the sky-blue'. When the Reynolds
number increases, the oscillatory instability of flow regime J is
observed.

Another characteristic of the flow regimes V and J is the precession
speed in the azimuthal direction $w\,=\,\omega/(2\pi)$. Dependence
of precession speed on the Reynolds number for two values of $b$ is
presented in Fig.~\ref{fig3}.

\subsection{Flow with a precessing vortex (regime V)}

\begin{figure}[H]
1 \includegraphics[width=3.8cm,height=3.8cm,angle=-90]{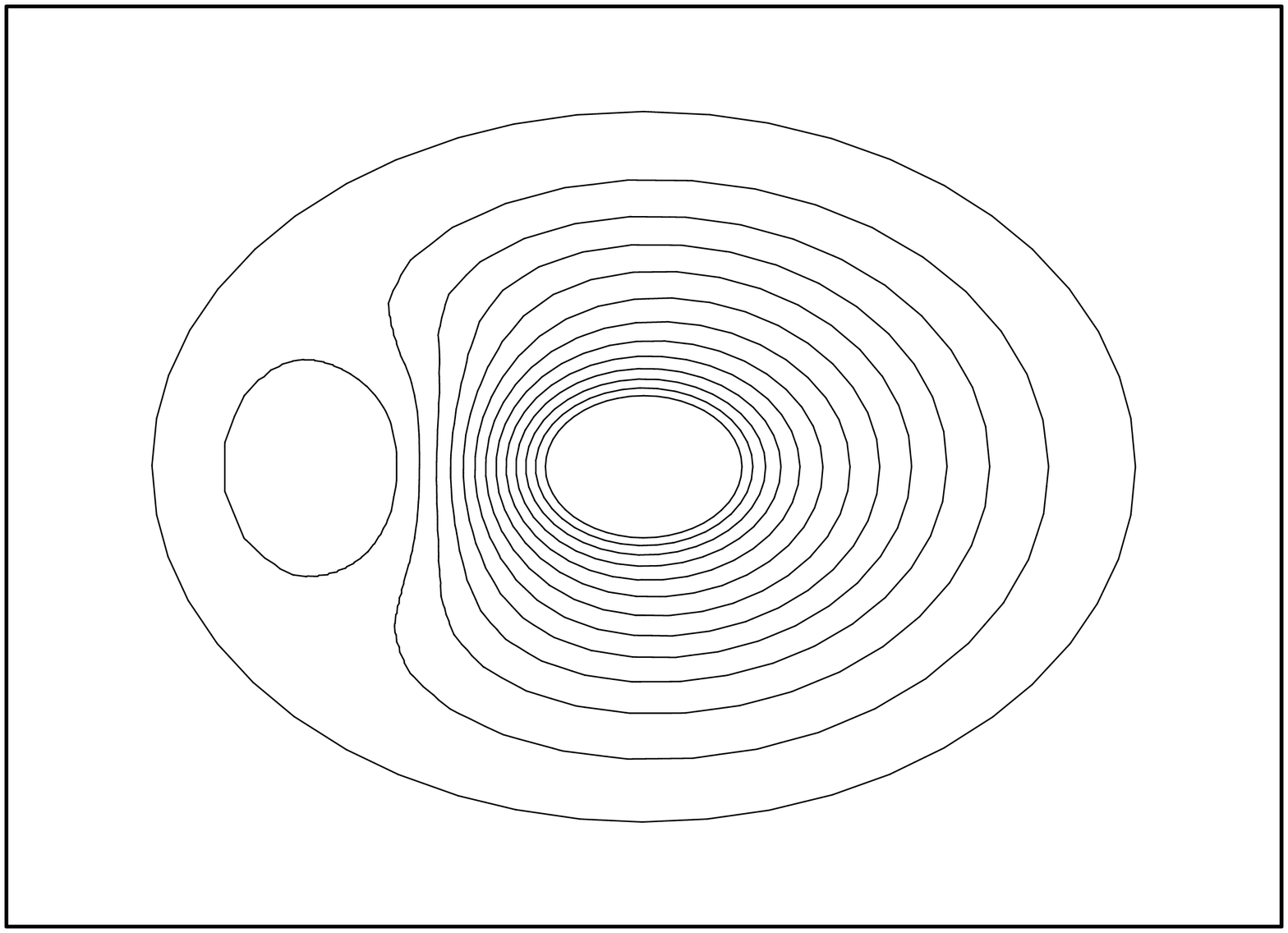} 2
\includegraphics[width=3.8cm,height=3.8cm,angle=-90]{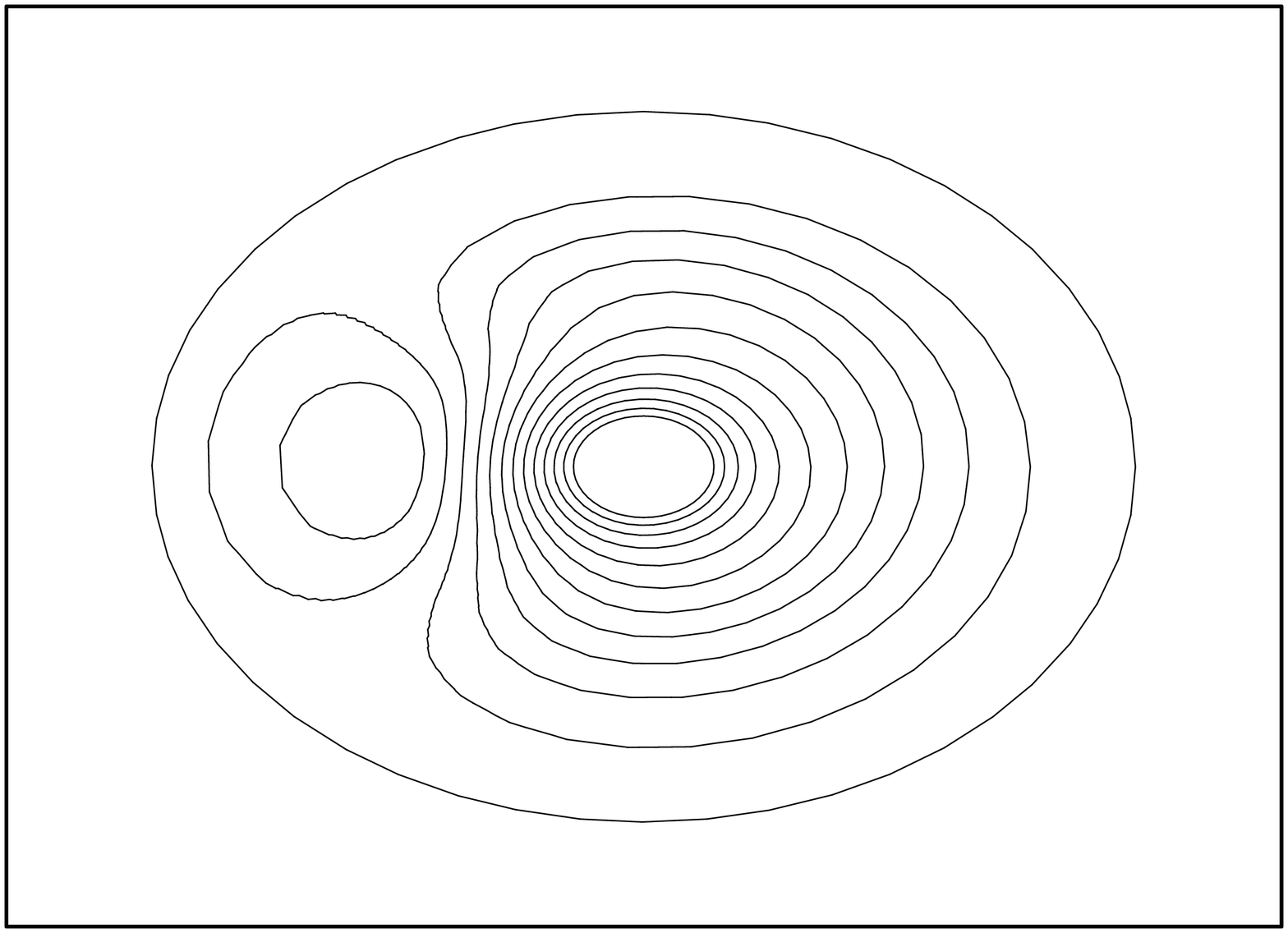}
3 \includegraphics[width=3.8cm,height=3.8cm,angle=-90]{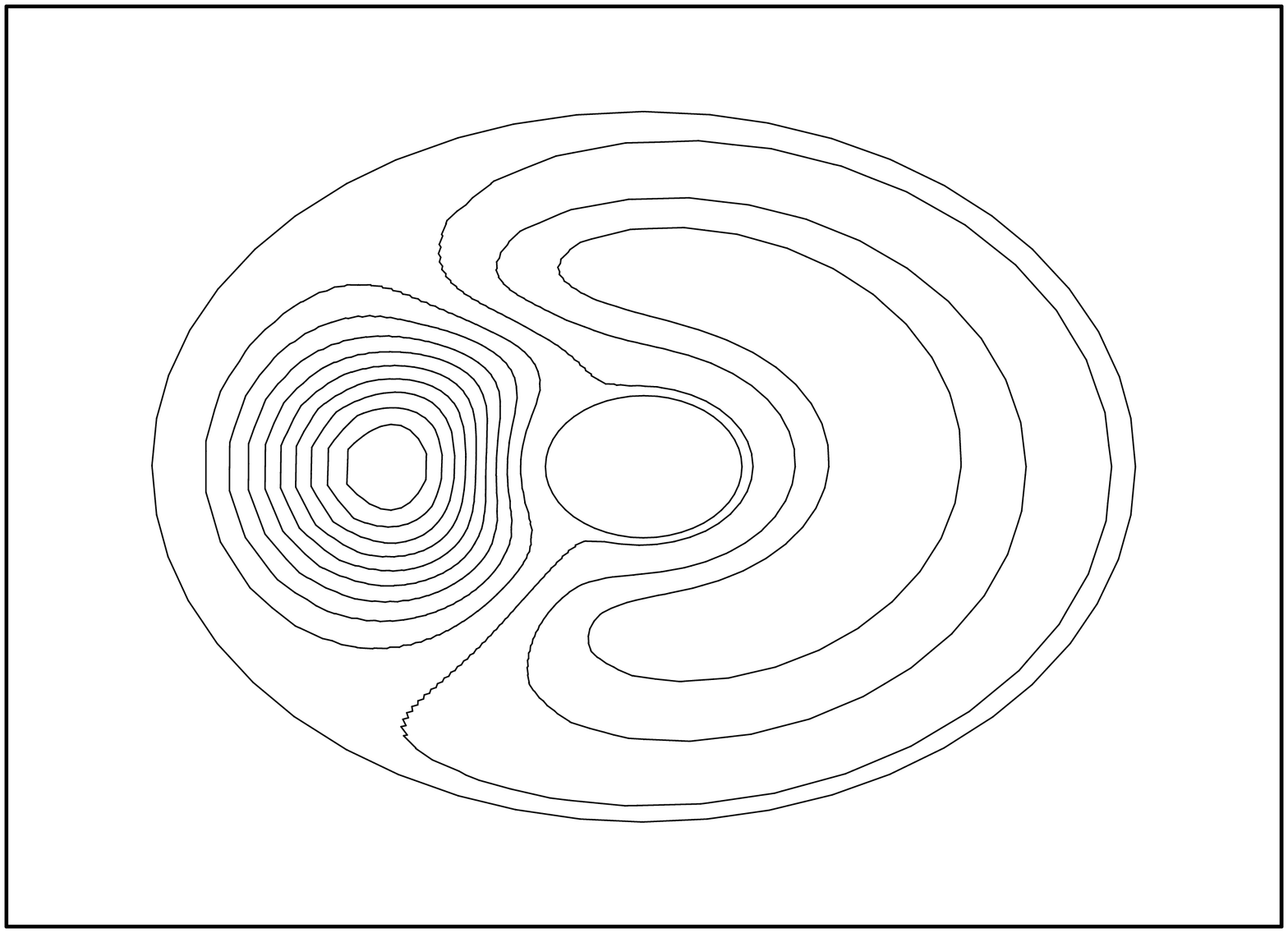} 4
\includegraphics[width=3.8cm,height=3.8cm,angle=-90]{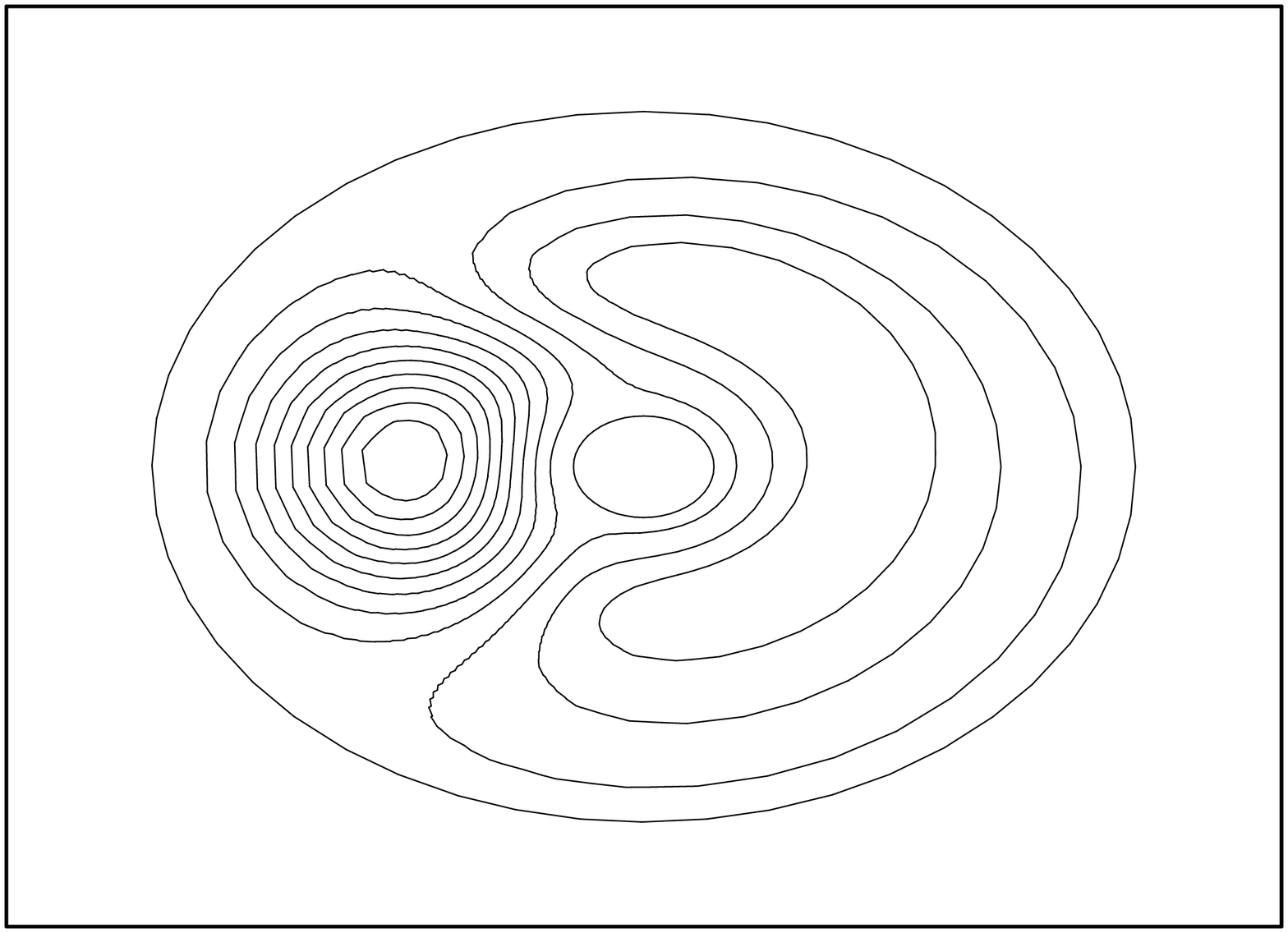}
\caption{Regime V, streamlines (1, 2) and level lines of the stream
function  perturbation $\psi(r,\theta)$ (3, 4); 1 and 3 correspond
to $b=5$, $\RE=320$; 2 and 4 correspond to $b=7$, $\RE=200$. }
 \label{fig4}
\end{figure}

\begin{figure}[H] \vspace{-2mm}
\centering
\includegraphics[width=6.6cm,angle=-90]{fig5a.eps}\hspace{26mm}
\includegraphics[width=4.8cm,angle=-90]{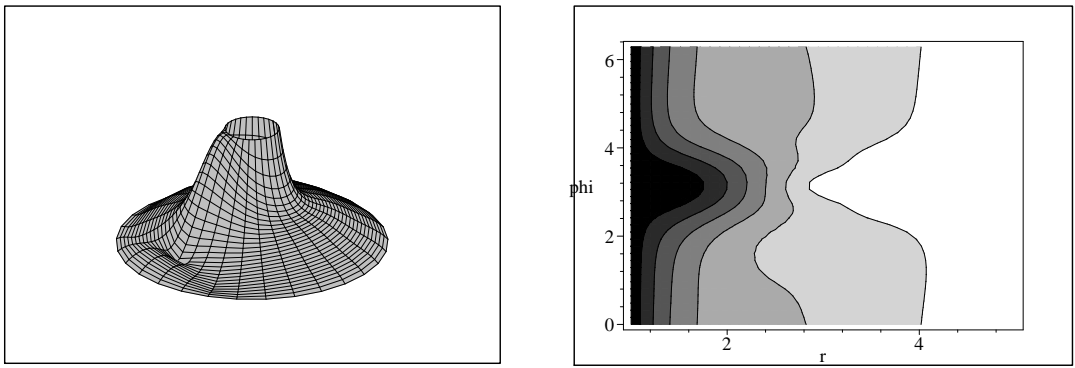}
\vspace{-16mm} \caption{Regime V, $b=5$, $\RE=320$. Left: graph of
the absolute velocity $|{\bf{v}}(r,\theta)|$. Right: contour plot of
the absolute velocity.}
 \label{fig5}
\end{figure}

\begin{figure}[H]
\centering
\includegraphics[width=4.8cm,angle=-90]{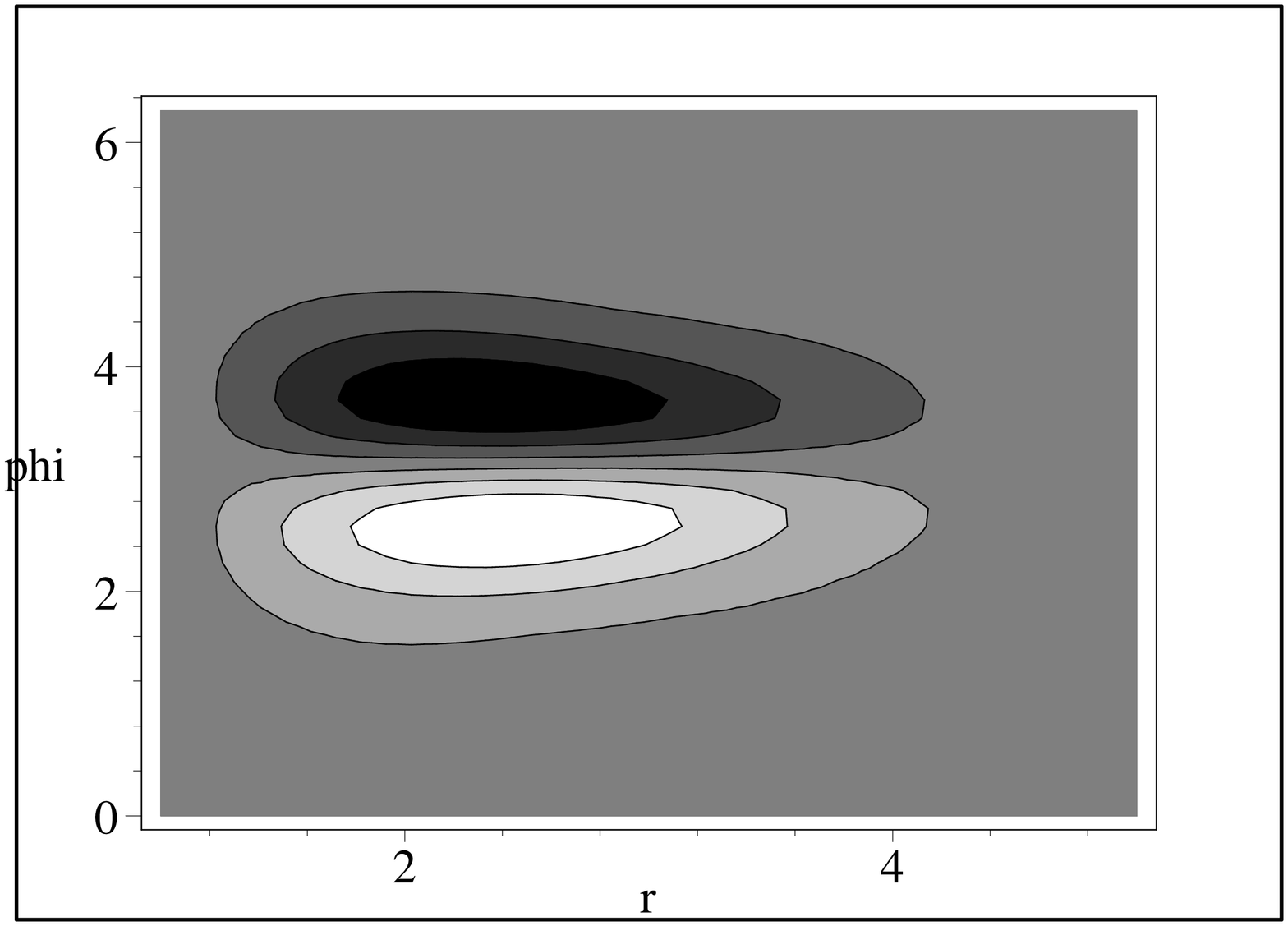}\hspace{8mm}
\includegraphics[width=4.8cm,angle=-90]{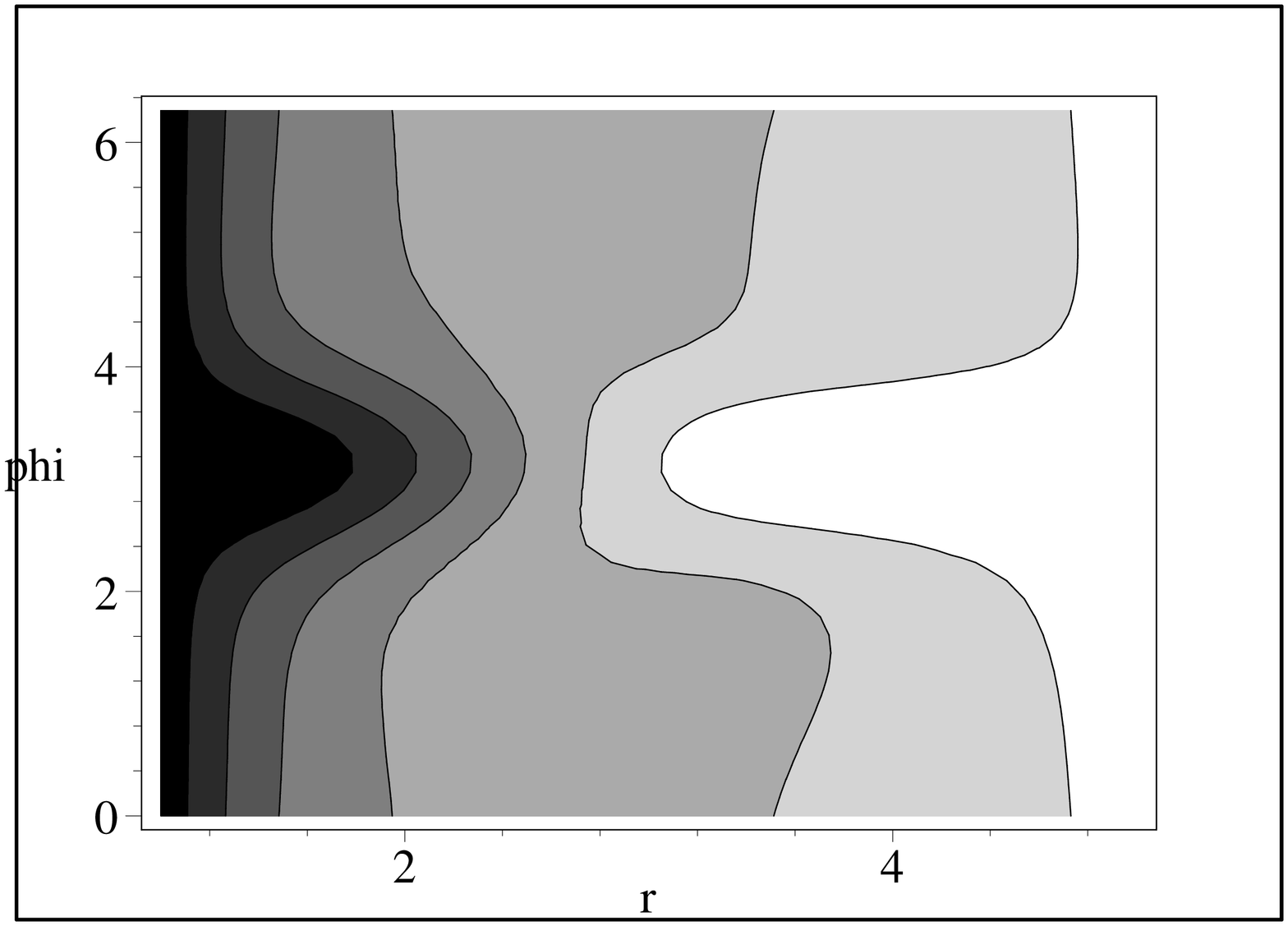}
 \caption{Contour plots of the radial and azimuthal velocity
$u(r,\theta)$, $v(r,\theta)$. Flow with precessing vortex; $b=5$,
$\RE=320$}
 \label{fig6}
\end{figure}

One of the flows observed in numerical experiments is the regime V:
periodic flow of the type of a traveling wave with the single vortex
that slowly precesses in the azimuthal direction (the direction of
the vortex precession coincides with the direction of the inner
circle rotation). Onset of this regime is linked to the loss of
stability of the base Couette flow. Regime V appears for both sets
of basis functions: a set of the eigenfunctions of the problem
(\ref{2.5})--(\ref{2.6}) and a set of the polynomials. Typical
results of numerical experiment for the flow of the type V are
presented in Figs.~\ref{fig4}--\ref{fig8}. Computation is performed
at $L=3$, $N=9$, basis composed of the eigenfunctions of
(\ref{2.5})--(\ref{2.6}) is used.

The streamlines and the level lines of the stream function
perturbation are plotted in Fig.~\ref{fig4} (location of the
soliton-like vortex is clearly visible). Graph of the absolute
velocity $|{\bf{v}}(r,\theta)|$  for the flow with the single vortex
(regime V) is exposed in Fig.~\ref{fig5} (left). Contour plot of the
absolute velocity in the rectangle $r \in [1,b]$, $\theta \in [0,2
\pi]$ is shown in Fig.~\ref{fig5} (right). Regions of the different
flow intensity are painted with different colors. The value
$|\bf{v}|$ is taken as the measure of the flow intensity. The
highest flow intensity is observed in the area close to the inner
boundary (black color). Note that the values of fluid velocity at the
inner and outer boundaries are $|{\bf{v}}|=1$ and $|{\bf{v}}|=0$ correspondingly, and
$|{\bf{v}}|=k/7$ $(k=1, 2,...,6)$ for level lines in Fig.5 (right).
The contour plots of radial velocity $u(r,\theta)$ and azimuthal velocity
$v(r,\theta)$ are presented in Fig.~\ref{fig6}.
\vspace{-2mm}
\begin{figure}[H]
\hspace{12mm}
\includegraphics[width=6.8cm,angle=-90]{fig7a.eps}\hspace{26mm}
\includegraphics[width=6.8cm,angle=-90]{fig7b.eps}
\vspace{-15mm} \caption{Dependence of radial velocity $u(r,\theta)$
and of azimuthal velocity $v(r,\theta)$ on $\theta$ at
$r=1+(3k-1)(b-1)/20$, where $k$ is the number of the line. Flow with
precessing vortex.}
 \label{fig7}
\end{figure}
\vspace{-2mm}
\begin{figure}[H]
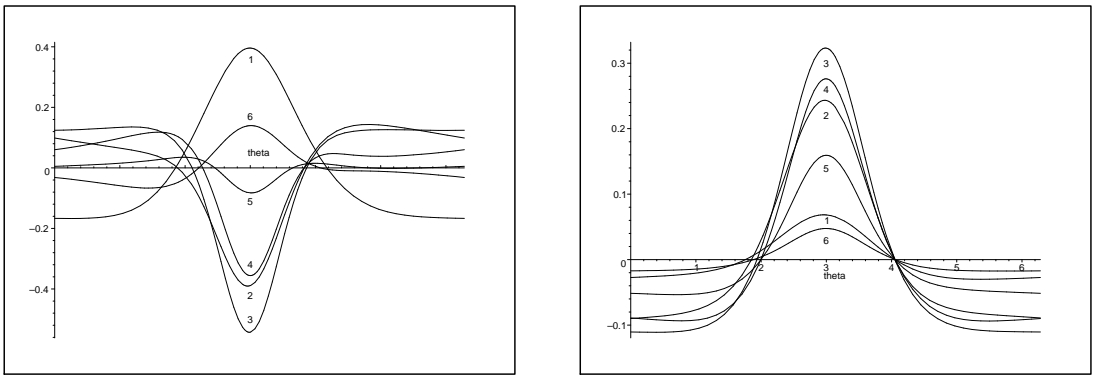

\hspace{12mm}
\includegraphics[width=6.8cm,angle=-90]{fig8a.eps}\hspace{26mm}
\includegraphics[width=6.8cm,angle=-90]{fig8b.eps}
\vspace{-16mm} \caption{Dependence of vorticity
$\Delta\psi(r,\theta)$ and of stream function perturbation
$\psi(r,\theta)$  on $\theta$ at $r=1+(3k-1)(b-1)/20$, where $k$ is
the number of the line. Flow with precessing vortex.}
 \label{fig8}
\end{figure}\vspace{-2mm}

In Figs.~\ref{fig7}--\ref{fig8} radial and azimuthal velocity
profiles, vorticity profiles, and stream function perturbation
profiles in the cross-sections $r=const$ are presented. Note that
zeros of the radial velocity $u(r,\theta)$ almost coincide with the
extreme points of the azimuthal velocity $v(r,\theta)$, and zeros of
$\Delta\psi(r,\theta)$ almost coincide with zeros of
$\psi(r,\theta)$. Extreme points of $\Delta\psi(r,\theta)$ and
extreme points of $\psi(r,\theta)$ also almost coincide.

\subsection{Flow with a precessing jet (regime J)}

Apart from the regime with precessing vortex, another type of
stationary flow is observed in numerical experiments. It is the
slowly precessing in azimuthal direction stationary structure, whose
main feature is existence of the jet-like area of intensive fluid
motion.

\begin{figure}[H]
1 \includegraphics[width=3.8cm,height=3.8cm,angle=-90]{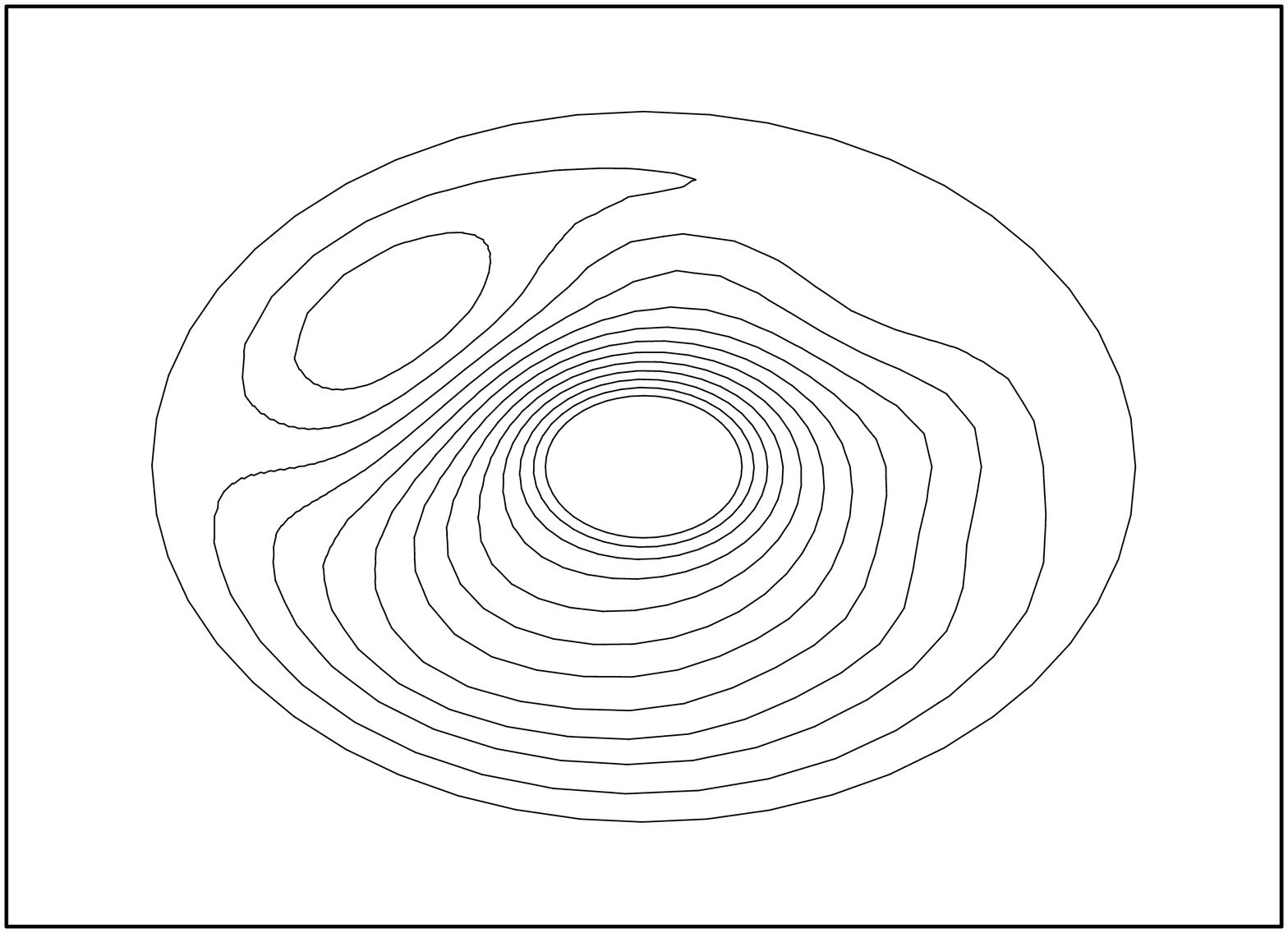} 2
\includegraphics[width=3.8cm,height=3.8cm,angle=-90]{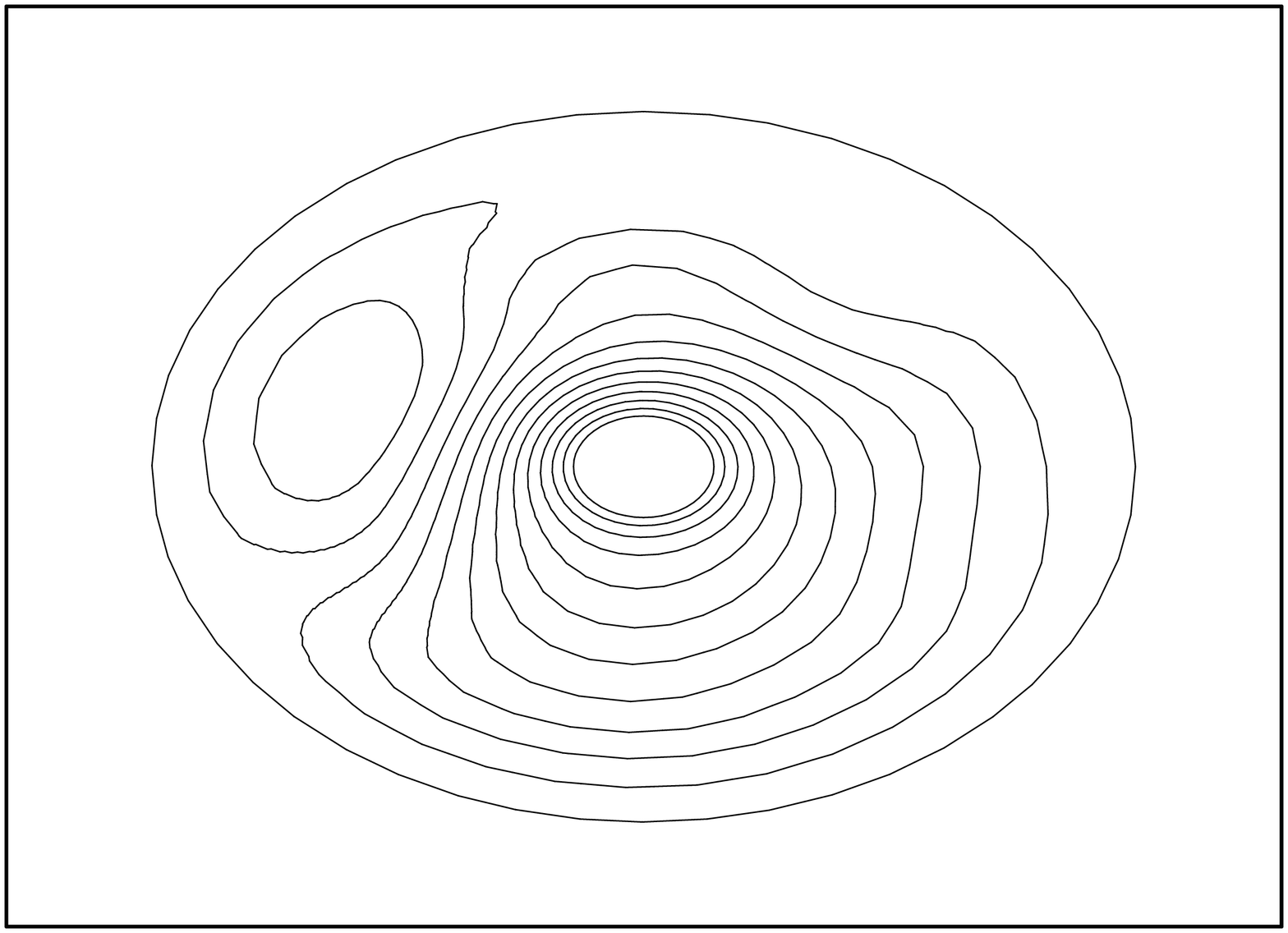}
3 \includegraphics[width=3.8cm,height=3.8cm,angle=-90]{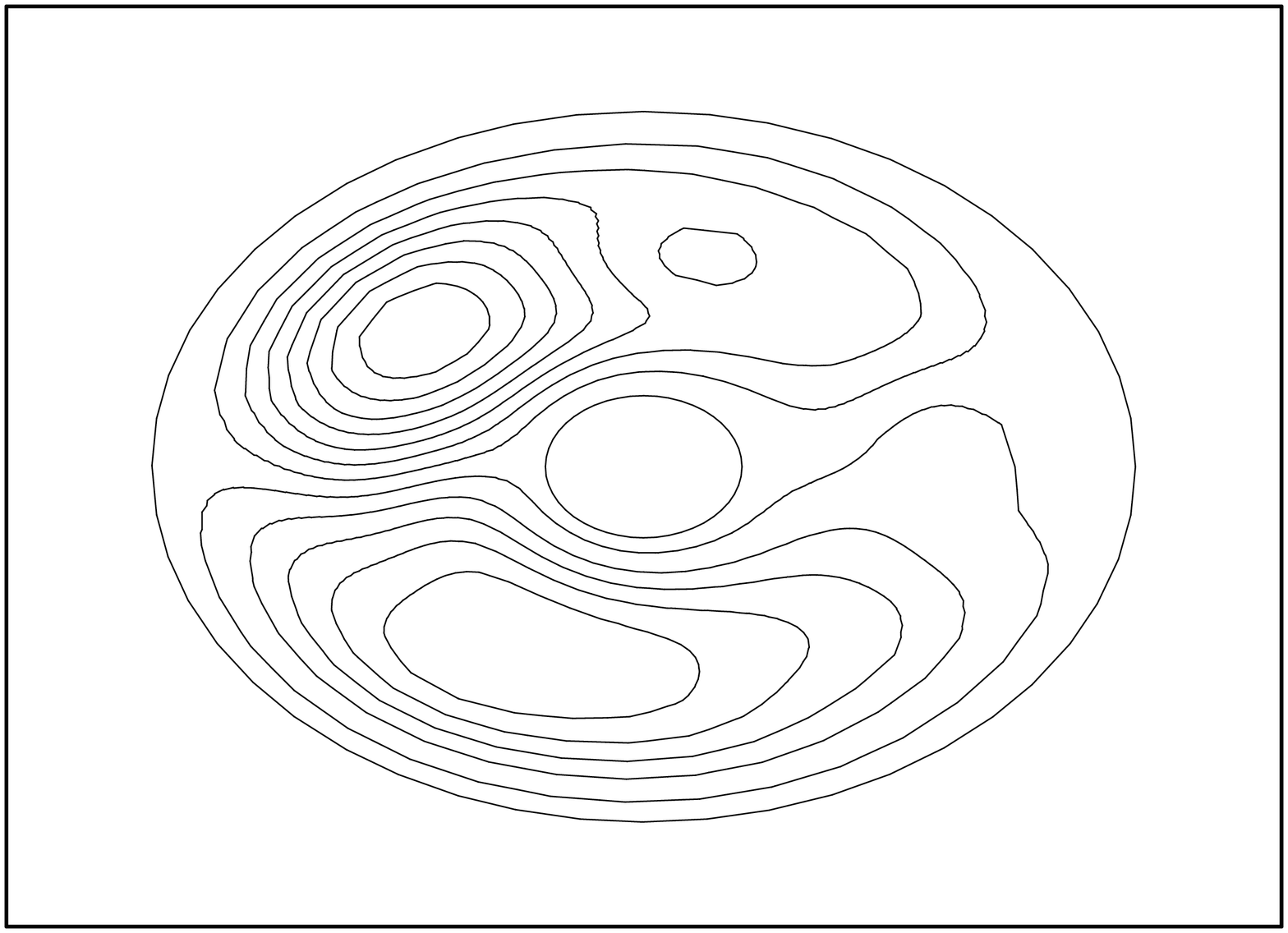} 4
\includegraphics[width=3.8cm,height=3.8cm,angle=-90]{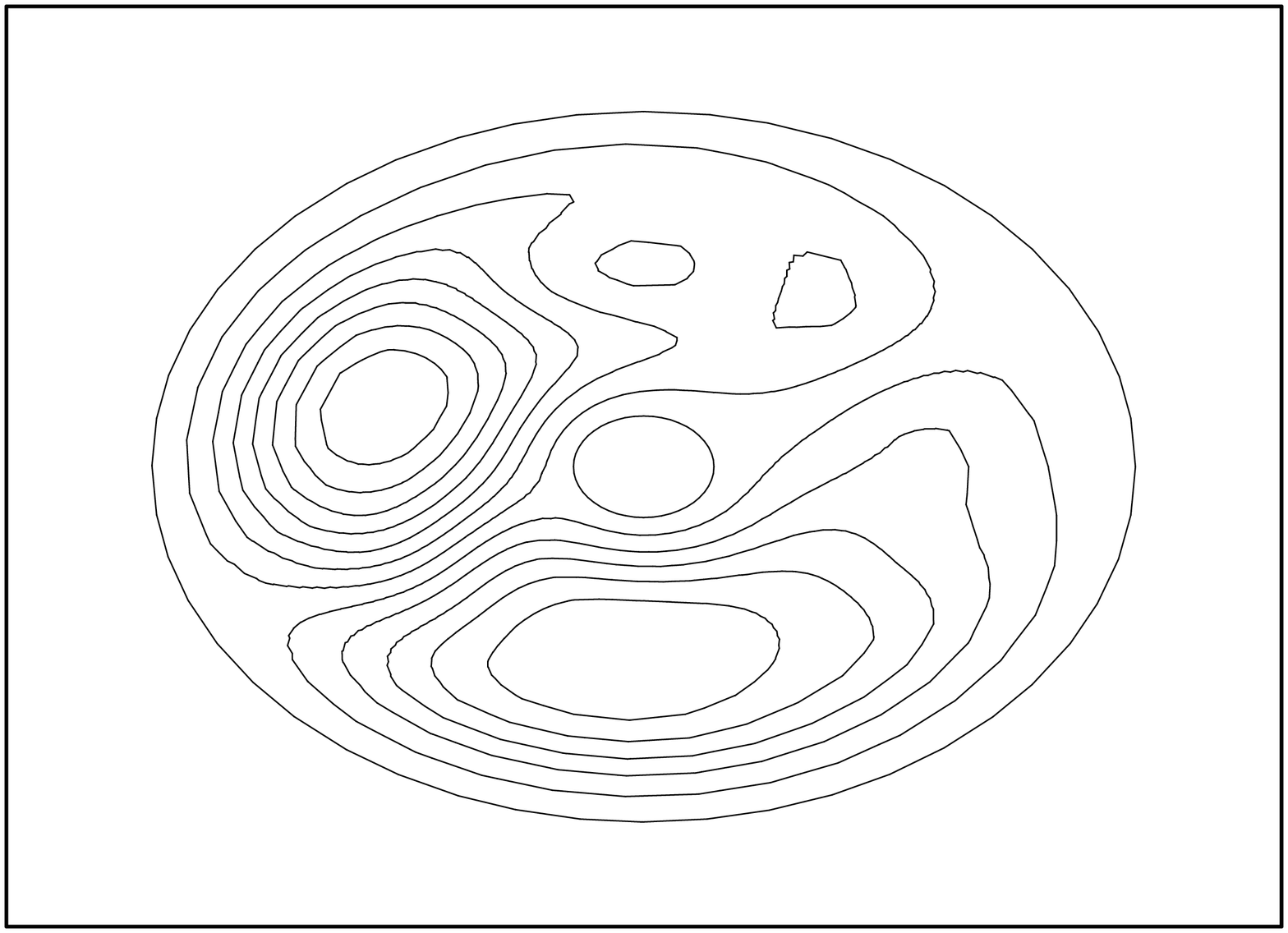}
\caption{Regime J, streamlines (1, 2) and level lines of the stream
function  perturbation $\psi(r,\theta)$ (3, 4); 1 and 3 correspond
to $b=5$, $\RE=320$; 2 and 4 correspond to $b=7$, $\RE=200$. Basis
composed of the eigenfunctions of (\ref{2.5})--(\ref{2.6}) is used
in both cases.}
 \label{fig9}
\end{figure}

\begin{figure}[H]
\centering
\includegraphics[width=6.6cm,angle=-90]{fig10a.eps}\hspace{26mm}
\includegraphics[width=4.8cm,angle=-90]{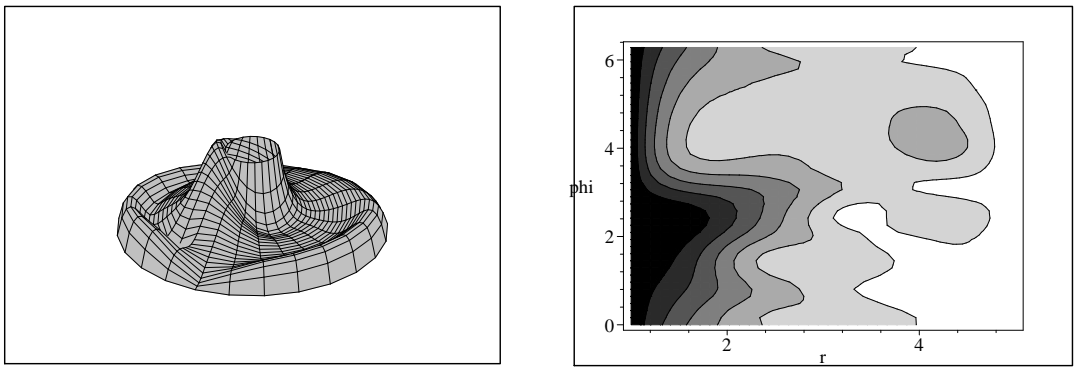}
\vspace{-16mm} \caption{Regime J, $b=5$, $\RE=320$. Left: graph of
the absolute velocity $|{\bf{v}}(r,\theta)|$. Right: contour plot of
the absolute velocity.}
 \label{fig10}
\end{figure}

\begin{figure}[H]
\centering
\includegraphics[width=4.8cm,angle=-90]{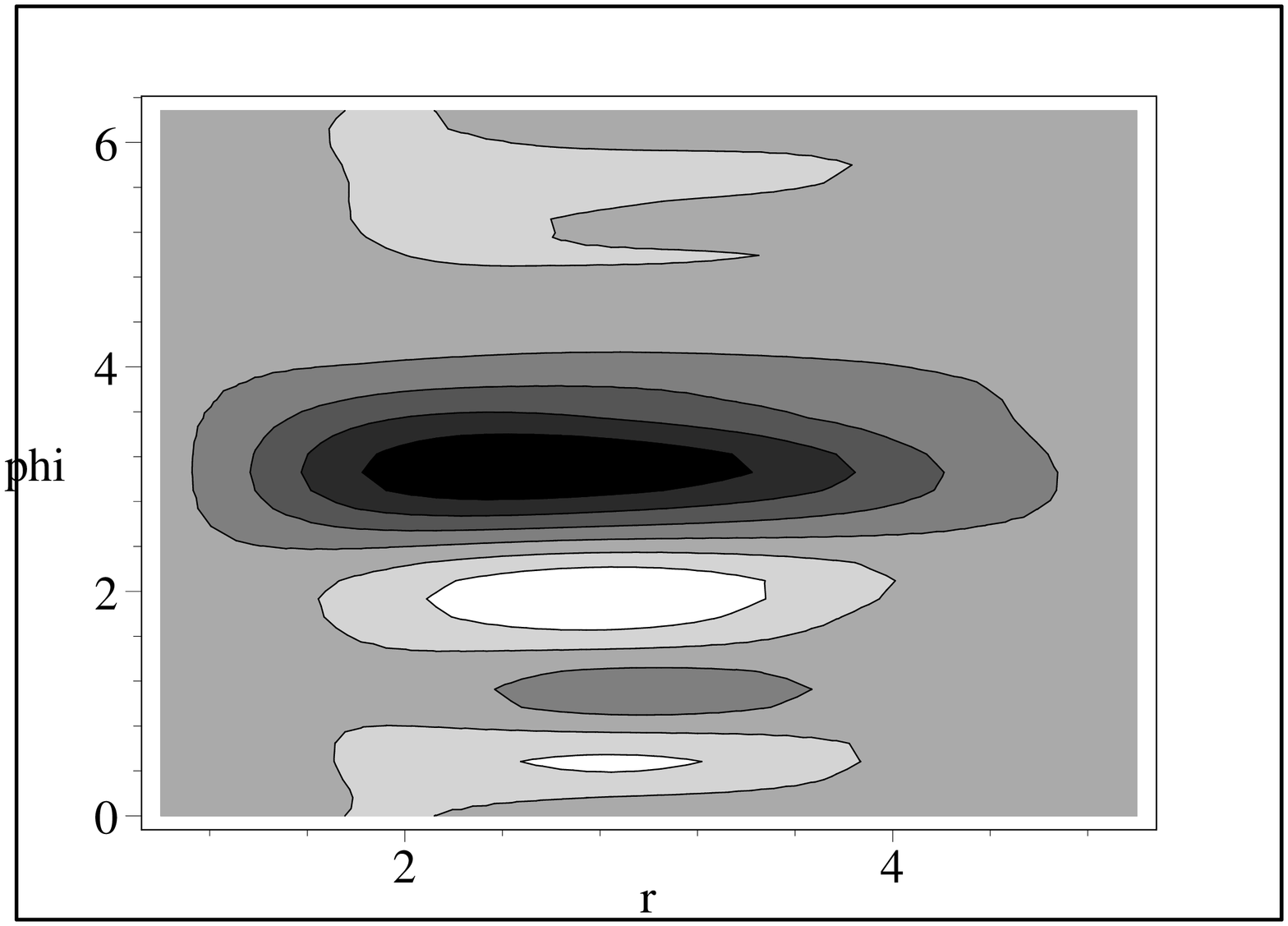} \hspace{8mm}
\includegraphics[width=4.8cm,angle=-90]{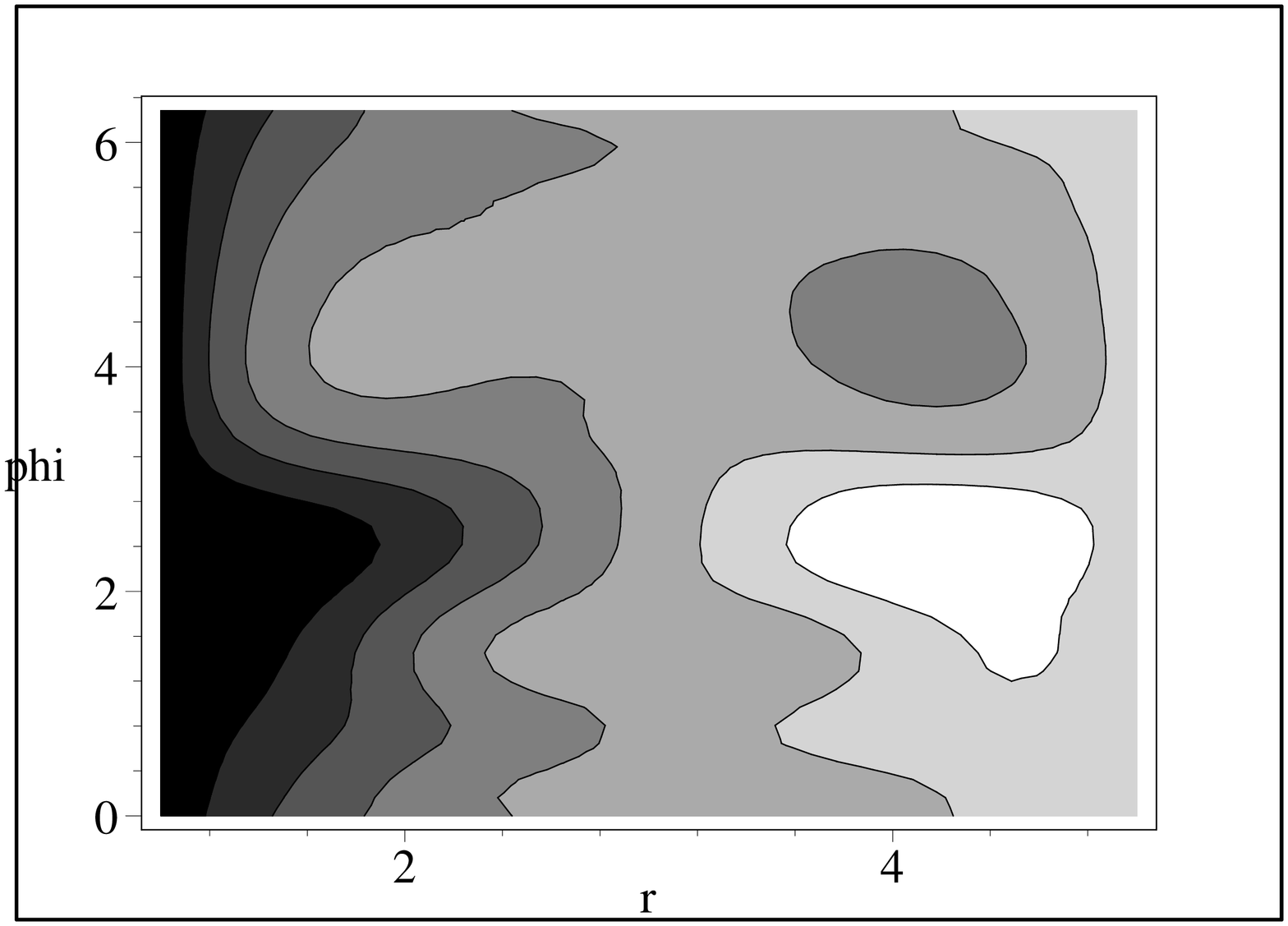}
\caption{Contour plot of the radial velocity $u(r,\theta)$ and of
the azimuthal velocity $v(r,\theta)$. Flow with precessing jet;
$b=5$, $\RE=320$.}
 \label{fig11}
\end{figure}

\begin{figure}[H]
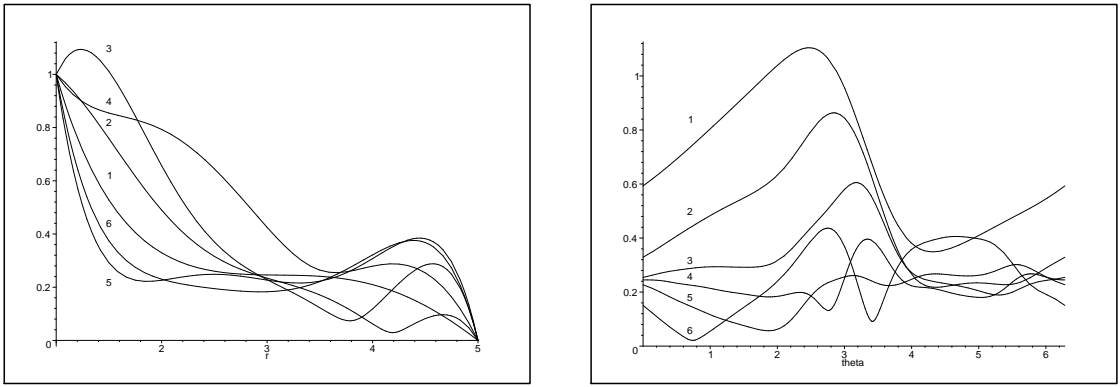

\hspace{12mm}
\includegraphics[width=7cm,angle=-90]{fig12a.eps}\hspace{26mm}
\includegraphics[width=7cm,angle=-90]{fig12b.eps}
\vspace{-15mm} \caption{Left: radial profiles of the absolute
velocity $|{\bf{v}}(r,\theta)|$ at $\theta=(k-1)\pi/3$, where $k$ is
the number of the line. Right: azimuthal profiles of
$|{\bf{v}}(r,\theta)|$ at $r=1+(3k-1)(b-1)/20$, where $k$ is the
number of the line. Flow with precessing jet.}
 \label{fig12}
\end{figure}

Results of numerical experiments for regime J are presented in
Figs.~\ref{fig9}--\ref{fig14}. The streamlines and the level lines
of the stream function perturbation are plotted in Fig.~\ref{fig9}.
There is also the precessing vortex, but its shape is more
complicated than that for the regime V. The vortex is stretched in
the azimuthal direction.

Graph of the absolute velocity $|{\bf{v}}(r,\theta)|$  for the flow
with precessing jet (regime J) and contour plot of the absolute
velocity in the rectangle $r \in [1,b]$, $\theta \in [0,2 \pi]$ are
shown in Fig.~\ref{fig10} (as before, $|\bf{v}|$ is taken as the
measure of the flow intensity). The most intensive flow is observed
in the certain region close to the inner boundary. Comparison with
Fig.~\ref{fig5} shows that the boundary layer near the inner
boundary have qualitatively different form for regimes V and J.

In Figs.~\ref{fig10}--\ref{fig14} various features of the flow
regime J are illustrated ($b=5$, $\RE=320$). The contour plots of
radial velocity $u(r,\theta)$ and azimuthal velocity $v(r,\theta)$
are presented in Fig.~\ref{fig11}. The jet-like area of intensive
fluid motion is clearly visible in Fig.~\ref{fig11} (left).

Radial profiles of the absolute velocity $|{\bf{v}}(r,\theta)|$ for
fixed $r$ are plotted in Fig.~\ref{fig12} (left), azimuthal profiles
of the absolute velocity for fixed $\theta$ are presented in
Fig.~\ref{fig12} (right). Azimuthal profiles of velocities
$u(r,\theta)$ and $v(r,\theta)$, profiles of vorticity $\Delta
\psi(r,\theta)$ and profiles of the stream function perturbation
$\psi(r,\theta)$ for $r\,=\,const$ are plotted in
Figs.~\ref{fig13}--\ref{fig14}.

Profiles of azimuthal velocity for constant $r$ chosen near to the
inner or outer circle have the shock-wave type. The jump from larger
to smaller values of azimuthal velocity in the direction of
precession is observed near the inner circle while the jump from
smaller to larger azimuthal velocities is observed near the outer
circle. Profiles of radial velocity have the soliton-like shape for
all fixed values of $r$. Note that zeros of the radial velocity
$u(r,\theta)$ almost coincide with extreme points of the azimuthal
velocity $v(r,\theta)$.
\begin{figure}[H]
\hspace{12mm}
\includegraphics[width=7cm,angle=-90]{fig13a.eps}\hspace{26mm}
\includegraphics[width=7cm,angle=-90]{fig13b.eps}
\vspace{-15mm} \caption{Dependence of radial velocity $u(r,\theta)$
and of azimuthal velocity $v(r,\theta)$ on $\theta$ at
$r=1+(3k-1)(b-1)/20$, where $k$ is the number of the line. Flow with
precessing jet.}
 \label{fig13}
\end{figure}
\begin{figure}[H]
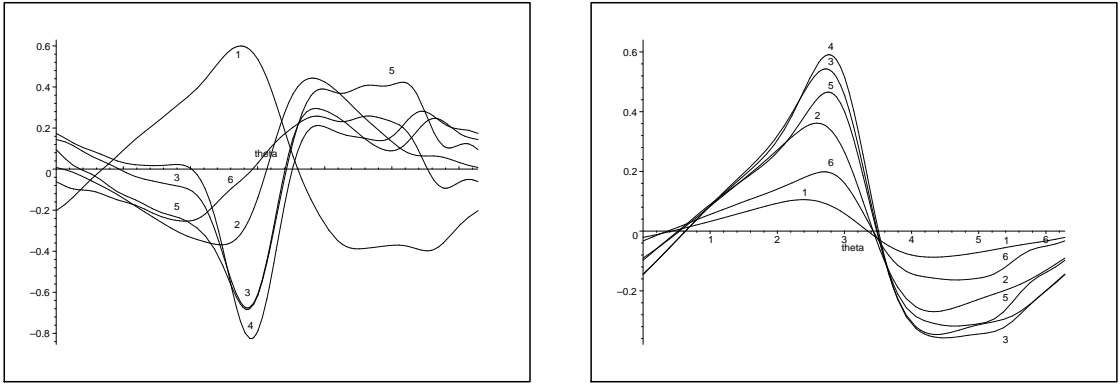

\hspace{12mm}
\includegraphics[width=7cm,angle=-90]{fig14a.eps}\hspace{26mm}
\includegraphics[width=7cm,angle=-90]{fig14b.eps}
\vspace{-15mm} \caption{Dependence of vorticity
$\Delta\psi(r,\theta)$ and of stream function perturbation
$\psi(r,\theta)$  on $\theta$ at $r=1+(3k-1)(b-1)/20$, where $k$ is
the number of the line.  Flow with precessing jet.}
 \label{fig14}
\end{figure}

\begin{figure}[H]
 \hspace{8mm}
\includegraphics[width=7cm,height=6.4cm]{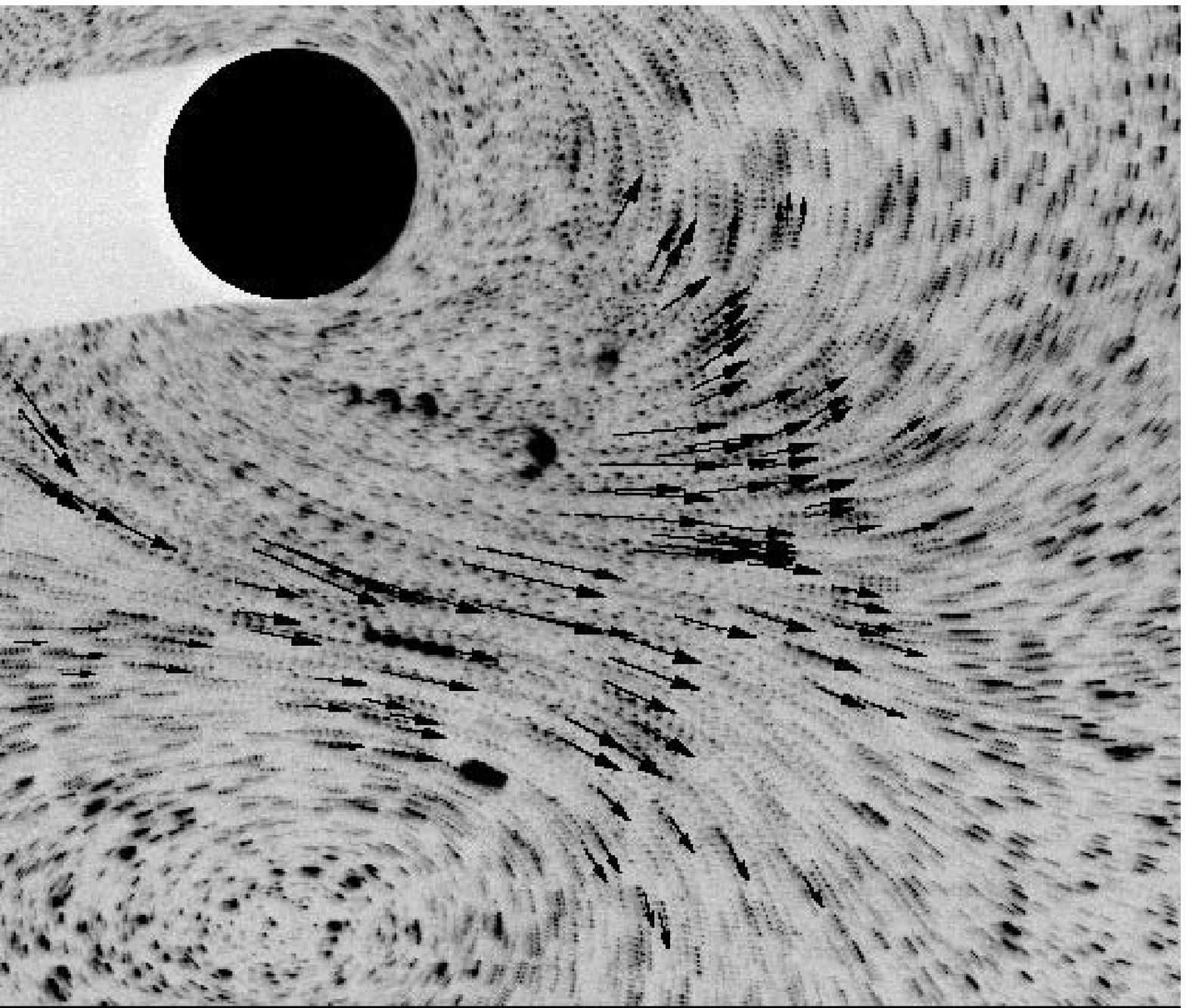} \hspace{8mm}
\includegraphics[width=7cm,height=6.4cm]{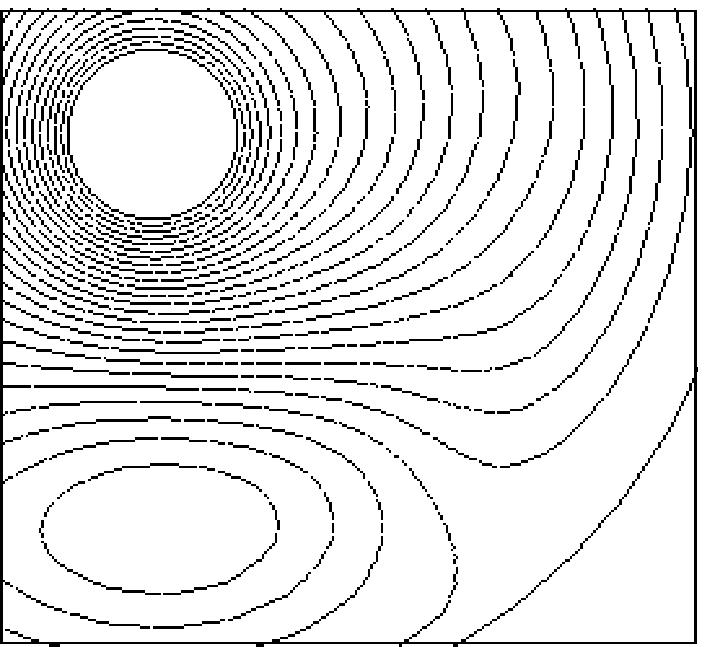}
\vspace{2mm} \caption{Left: experiment. Free upper surface.
Water-glycerol mixture, axial gap thickness $16 \cdot 10^{-3}\,m$,
radius of the cylinder $10^{-2}\,m$, outer radius of domain $23.2
\cdot 10^{-2}\,m$, kinematic viscosity $0.43 \cdot 10^{-4}\,m^2/s$,
rotation speed of the inner cylinder $230\,rpm$, $\RE \sim 56$.
Right: Regime J, streamlines, $b=5$, $\RE=320$.}
 \label{figv}
\end{figure}

It is interesting to compare results of computation with results of
the experiments in which the jet-like flow regimes are observed. In
Fig.~\ref{figv} (published with permission of V.\,A.\,Vladimirov and
P.\,V.\,Denissenko), visualization of the experiment with the
jet-like structure is presented ($\RE \sim 56$). The qualitative
similarity of the experimental results (Fig.~\ref{figv}, left) and
of the numerical results (Fig.~\ref{figv}, right) is obvious. We
repeat once again that the quantitative comparison can not be
performed because the physical experiment is essentially
three-dimensional while the computation considered in this paper is
two-dimensional.

\vspace{-2mm}
\subsection*{Acknowledgements}

This research is partially supported by the Russian Ministry of
Education (programme 'Development of the research potential of the
high school', grant 2.1.1/554), by Russian Foundation for Basic
Research (grants 07-01-00389, 08-01-00895, and 07-01-92213 NCNIL),
and by SRDF grant RUM1-2842-RO-06. This work was done in the
framework of European Research Group 'Regular and Chaotic
Hydrodynamics'.

\end{document}